\documentclass{article}

\newcommand{\mf}{\mathbf{f}}
\newcommand{\mpsi}{\boldsymbol{\psi}}

\newcommand{\mB}{\boldsymbol{B}}
\newcommand{\mY}{\boldsymbol{Y}}
\newcommand{\mZ}{\boldsymbol{Z}}
\newcommand{\mV}{\boldsymbol{V}}
\newcommand{\mW}{\boldsymbol{W}}

\newcommand{\mD}{\mathbb{D}}

\newcommand{\p}[1]{(\ref{#1})}

\newcommand{\cF}{{\cal F}}

\newcommand{\cD}{{\cal D}}

\newcommand{\cG}{{\cal G}}
\newcommand{\cH}{{\cal H}}

\newcommand{\cE}{{\cal E}}
\newcommand{\cW}{{\cal W}}

\newcommand{\cL}{{\cal L}}

\newcommand{\cbW}{\overline{\cal W}}

\newcommand{\bD}{{\overline D}}

\newcommand{\bW}{{\overline W}}

\newcommand{\bxi}{{\bar\xi}}

\newcommand{\bpsi}{{\bar\psi}{}}

\newcommand{\uA}{{\underline A}}
\newcommand{\uB}{{\underline B}}
\newcommand{\uC}{{\underline C}}
\newcommand{\uD}{{\underline D}}

\newcommand{\ucD}{{\underline {\cal D}}}
\newcommand{\ucE}{{\underline {\cal E}}}
\newcommand{\ualpha}{{\underline \alpha}}
\newcommand{\ubeta}{{\underline \beta}}
\newcommand{\ulambda}{{\underline \lambda}}

\newcommand{\be}{\begin{equation}}
\newcommand{\ee}{\end{equation}}
\newcommand{\bea}{\begin{eqnarray}}
\newcommand{\eea}{\end{eqnarray}}

\newcommand{\ba}{\begin{array}} \newcommand{\ea}{\end{array}}

\def\im{{\rm i}}
\newcommand{\tr}{{\rm Tr}}

\newcommand{\nn}{\nonumber}

\usepackage{amscd,amsmath,amssymb}

\begin{document}

\thispagestyle{empty}
\vspace{2cm}
\begin{center}
{\Large\bf Component $d=6$ Born-Infeld theory } \\
\vspace{0.3cm}

{\Large\bf  with $N=(2,0)\rightarrow N=(1,0)$ supersymmetry breaking}
\end{center}
\vspace{2cm}

\begin{center}
{\large\bf  N. Kozyrev${}^{a}$
}
\end{center}

\begin{center}
${}^a$ {\it Bogoliubov  Laboratory of Theoretical Physics, JINR, 141980 Dubna,
Russia}
\end{center}

\begin{center}
{\tt nkozyrev@theor.jinr.ru}
\end{center}
\vspace{2cm}

\begin{abstract}
\noindent
The formalism of nonlinear realizations is used to construct a theory with $1/2$ partial breaking of global supersymmetry with the $N=(1,0)$, $d=6$ abelian vector multiplet as a Goldstone superfield. Much like the case of the $N=2$, $d=4$ Born-Infeld theory, proper irreducibility conditions of the multiplet are selected by the invariance with respect to the external automorphisms of the Poincar\'e superalgebra. They are found in the lowest nontrivial order in the auxiliary field. The fermionic contributions to the Bianchi identity are restored by assuming its covariance with respect to broken supersymmetry. The invariance of the action with respect to unbroken supersymmetry is checked in the lowest order in the fermionic fields. The supersymmetry preserving reduction of the $d=6$ action to four dimensions is performed, resulting in the $N=4$, $d=4$ Born-Infeld theory. As expected, the reduced action enjoys $U(1)$ self-duality.

\end{abstract}

\newpage
\setcounter{page}{1}
\setcounter{equation}{0}

\section*{Introduction}

A lot of work has already been devoted to the study of the supersymmetric Born-Infeld theories. In the string theories, they arise in the effective description of the $D$-branes \cite{tsey1}. In the supersymmetric context, they appear while studying of partial spontaneous breaking of global supersymmetry with vector multiplets as Goldstone superfields.  One of the simplest and most familiar systems of this kind is the theory of the $N=1$, $d=4$ vector multiplet with additional spontaneously broken $N=1$, $d=4$ supersymmetry which is a direct generalization of the original Born-Infeld theory \cite{origBI}. Its superfield action was constucted by Ceccotti and Ferrara \cite{1superBI}. The fact that this action is invariant with respect to additional spontaneously broken supersymmetry was established by Bagger and Galperin \cite{BGvect}. They obtained the superfield Lagrangian as a composite $N=1$, $d=4$ superfield which, together with the Goldstone fermion, provides the realization of the $N=2$, $d=4$ supersymmetry. Also, they proved self-duality of this action with respect to the Legendre transformations. After that many other ways to construct the $N=2$, $d=4$ Born-Infeld theory were found such as nilpotent superfields \cite{nilpBI} and superembedding approach \cite{superembBI}. It was found how to explicitly construct the component action of this theory using the formalism of nonlinear realizations \cite{compN2d4BI}.

One may also try to construct analogous theories with higher supersymmetry breaking the $N=4$, $d=4$ supersymmetry with $N=2$, $d=4$  or $N=(2,0)$, $d=6$ supersymmetry with $N=(1,0)$, $d=6$ vector multiplets as Goldstone superfields. The first of these theories can be produced by the dimensional reduction of the second, and they describe $D3$- and $D5$-branes in $D=6$, respectively. The attempts to find their superfield actions, however, were not as successful as with the $N=2$, $d=4$ Born-Infeld theory. One superfield action was proposed in \cite{ketov1}. It satisfied the nonlinear constraint, later called the Ketov equation, which was a generalization of the constraint the Cecotti and Ferrara Lagrangian satisfied. However, it was criticized in  \cite{nlselfdual} as it is not possible to write down such a shift and broken supersymmetry transformations of the Goldstone bosons that are compatible with the mentioned constraint. Later analysis revealed \cite{sfN4BI1} that it is indeed possible to realize additional broken $N=2$, $d=4$ supersymmetry but only on the infinite set of $N=2$, $d=4$ superfields, satisfying an infinite number of constraints, and only one of these superfields is the proper superfield Lagrangian. A few terms in the power series expansion of the action were found this way, not contradicting those obtained in \cite{nlselfdual} from the requirement of self-duality and invariance with respect to shifts of the Goldstone bosons, but the computation of the whole action appeared to be possible only in principle. Even the exact solution of Ketov's condition (truncation of an infinite system of \cite{sfN4BI1}) appeared hard to find, with new terms appearing in the $20$th order in power expansion \cite{onroad}. The use of the formalism of nonlinear realizations allowed just to compute the equations of motion \cite{compN4d4BIold} and only in the specific limits. The computation of the action of the $N=(2,0)$, $d=6$ theory \cite{ketov1} also faced difficulties: it was argued \cite{nlselfdual} that the proposed action is not even $N=(1,0)$ supersymmetric and that it is not possible to write down the six dimensional action as an integral over the $N=(1,0)$, $d=6$ superspace or its supersymmetric subspaces.

Therefore, it would be reasonable to try to find an alternative way to deal with such theories. Indeed, one may try to make a theory treatable either by using more elaborate extensions of the superfield method, as was suggested in \cite{purespBI}, either by formulating the theory in terms of the component fields. In this paper, we consider the second option.

The component approach to the actions with partially spontaneously broken global supersymmetry was suggested in the papers \cite{cosetact}, \cite{partact} for three-dimensional theories with scalar fields and supersymmetric mechanics. It was used to construct the actions of the $N=2$, $d=3$ and $N=1$, $d=4$ chiral multiplets, as well as $N=2$, $d=4$ hypermultiplet, all with spontaneous breaking half supersymmetry \cite{scalaracts}. Later the component theories with the vector multiplets, $N=1$, $d=4$ \cite{compN2d4BI} and $N=2$, $d=3$ \cite{compN4d3BI}, were constructed. The basic point of the component approach is that it is possible to define the Goldstone fermionic superfield so that the broken supersymmetry is realized on this superfield and the spacetime coordinates in a very simple way, like in the work by Volkov and Akulov \cite{neugold}, while the $\theta$-coordinates of the superspace remain inert. If these conditions hold, the transformations of the first component of the Goldstone fermionic superfield mimic the transformations of the Volkov-Akulov fermion. Invariance of the action with respect to such transformations completely fixes its dependence on the Goldstone fermions. In particular, it implies that this fermion may only enter into the action only either through the matrix $\cE_A{}^B = \delta_A^B + \im(\bar\psi \gamma^B \partial_A \psi)$, which covariantizes the derivatives of all fields $\partial_A \rightarrow \cD_A = \big( \cE^{-1} \big)_A^B\partial_B$ and the integration measure $d^4 x \rightarrow d^4 x \det \cE$ \cite{neugold}, or the Wess-Zumino terms. The complete supersymmetric action, therefore, would be just a simple generalization of the bosonic action, and it would be only required to check its invariance with respect to the unbroken supersymmetry.

It should also be noted that of the two related Born-Infeld theories, $N=4$, $d=4$ and $N=(2,0)$, $d=6$, it makes more sense to construct the second one as the four dimensional theory could then be obtained by the dimensional reduction. Moreover, there are indications that actually the $N=(2,0)$, $d=6$ theory would be easier to construct. Indeed, one of the simplest theories with scalar and electromagnetic fields was analyzed in \cite{compN4d3BI}, where conclusion was reached that it would be highly desirable to formulate the irreducibility conditions of the multiplet in terms of the fermionic superfields. This would eliminate the necessity to solve nonlinear algebraic relations between derivatives of scalar fields and bosonic components of fermionic superfields, which appear in all theories with scalars and can be very complicated in the cases of high supersymmetry (examples can be found in \cite{scalaracts}). Also, the components of the vector multiplets, which correspond to the electromagnetic field, satisfy the differential identity (called the Bianchi identity). It should be derived as a consequence of the irreducibility conditions, and this is much easier to do if the conditions are formulated in terms of fermionic superfields. Also, in the theories with spontaneous breaking of supersymmetry this condition is typically highly nonlinear and should be proven equivalent to the usual $\partial_{[A}F_{BC]}=0$, which would also relate the true physical field strength $F_{AB}$ to the components of the multiplet. This is much simpler to do if the identity does not involve scalar fields. As the only physical boson of the $N=(1,0)$, $d=6$ multiplet is the electromagnetic field strength tensor $F_{AB}$, while $N=2$, $d=4$ supermultiplet has two additional scalars, the six dimensional case is preferable.

Therefore, our approach to construct the action is the following one.
\begin{itemize}
\item At first, we should derive proper irreducibility conditions of the $N=(1,0)$, $d=6$ vector multiplet from the assumption of covariance with respect to broken supersymmetry and the $SO(4)$ group (subgroup of the $SO(5)$ automorphisms of the $N=(2,0)$, $d=6$ superalgebra).
\item Secondly, as the consequence of the irreducibility conditions, the nonlinear Bianchi identities should be derived. Let us note that it is sufficient to find them in the bosonic limit and with the auxiliary field removed by its equation of motion. This is acceptable as we are going to construct the action without the auxiliary field, and the fermionic terms in the identity can be restored from the assumption of its covariance with respect to the broken supersymmetry. Then it should be shown that the found nonlinear identities are equivalent to the usual ones $\partial_{[A}F_{BC]}=0$. At the same time,  the expression of the physical bosonic field strength in terms of the bosonic components of the multiplet would be found.
\item Thirdly, the ansatz for the action should be constructed by covariantizing the well-known bosonic action with respect to broken supersymmetry and by adding the Wess-Zumino term. Finally, using the standard techniques, the transformation laws of the components with respect to unbroken supersymmetry should be derived and the invariance of the action proven in the lowest nontrivial approximation in the fermions.
\end{itemize}

\section{The superalgebra and the coset space}
The $N=(2,0)$, $d=6$ superalgebra is composed of two copies of $N=(1,0)$, $d=6$ superalgebras,
\be\label{N2d6susy}
\big\{ Q^i_\alpha, Q^j_\beta \big\}=2\epsilon^{ij}P_{\alpha\beta}, \;\;\; \big\{ S^i_\alpha, S^j_\beta \big\}=2\epsilon^{ij}P_{\alpha\beta},
\ee
as well the Lorentz algebra in $d=6$ and the $sp(2)\sim so(5)$ algebra of automorphisms. Indices $i,j =1,2$ are those of $SU(2)$ spinors, and $\alpha,\beta =1,\ldots,4$ are the indices of $so(1,5)\sim su^*(4)$ spinors. In this notation, $P_{\alpha\beta} =-P_{\beta\alpha}$ is the $d=6$ spacetime vector, $F_\alpha{}^\beta$ is the antisymmetric tensor if $F_\alpha{}^\alpha=0$, $C_{\alpha\beta} = C_{\beta\alpha}$ is the self-dual three-form, and so on.

The commutation relations of the $so(5)$  automorphism algebra in the basis with only one explicit $su(2)$ can be written as
\bea\label{so5coms}
&\big[ T^{ij}, T^{kl}  \big] = \im \big( \epsilon^{ik} T^{jl} + \epsilon^{jl}T^{ik} \big), \;\; \big[ T^{ij}, R^{kl}  \big] = \im \big( \epsilon^{ik} R^{jl} + \epsilon^{jl}R^{ik} \big), \;\; \big[ T^{ij},{\widetilde R}^{kl}  \big] = \im \big( \epsilon^{ik}{\widetilde R}^{jl} + \epsilon^{jl}{\widetilde R}^{ik} \big), &\nn \\
&\big[ R^{ij}, R^{kl}  \big] = \im \big( \epsilon^{ik} T^{jl} + \epsilon^{jl}T^{ik} \big), \;\; \big[ {\widetilde R}^{ij}, {\widetilde R}^{kl}  \big] = \im \big( \epsilon^{ik} T^{jl} + \epsilon^{jl}T^{ik} \big), \;\; \big[ R^{ij}, {\widetilde R}^{kl}  \big] = \im \big(\epsilon^{ik}\epsilon^{jl} + \epsilon^{jk}\epsilon^{il} \big)R_0,& \\
&\big[ R_0, R^{ij}\big] = \im {\widetilde R}^{ij},\;\; \big[ R_0, {\widetilde R}^{ij}\big] = -\im R^{ij}.&\nn
\eea
The generators of $so(5)$ commute with the supercharges as
\bea\label{so5Qcoms}
&\big[ T^{ij},Q^{k}_{\alpha}  \big]=\frac{\im}{2}\big( \epsilon^{ik}Q^j_\alpha + \epsilon^{jk}Q^i_\alpha  \big), \;\; \big[ T^{ij},S^{k}_{\alpha}  \big]=\frac{\im}{2}\big( \epsilon^{ik}S^j_\alpha + \epsilon^{jk}S^i_\alpha  \big), &\nn \\
&\big[ R^{ij},Q^{k}_{\alpha}  \big]=\frac{\im}{2}\big( \epsilon^{ik}S^j_\alpha + \epsilon^{jk}S^i_\alpha  \big), \;\; \big[ R^{ij},S^{k}_{\alpha}  \big]=\frac{\im}{2}\big( \epsilon^{ik}Q^j_\alpha + \epsilon^{jk}Q^i_\alpha  \big), & \\
&\big[ {\widetilde R}^{ij},Q^{k}_{\alpha}  \big]=\frac{\im}{2}\big( \epsilon^{ik}Q^j_\alpha + \epsilon^{jk}Q^i_\alpha  \big), \;\; \big[  {\widetilde R}^{ij},S^{k}_{\alpha}  \big]=-\frac{\im}{2}\big( \epsilon^{ik}S^j_\alpha + \epsilon^{jk}S^i_\alpha  \big),& \nn \\
&\big[ R_0, Q^i_\alpha  \big] =-\frac{\im}{2}S^i_\alpha, \;\;  \big[ R_0, S^i_\alpha  \big] =\frac{\im}{2} Q^i_\alpha.&\nn
\eea
For the purposes of the latter construction, only the generators $R^{ij}$ and $T^{ij}$, which form $so(4)$, are relevant.

The spontaneous breaking of half the supersymmetry can be achieved with the following coset element:
\be\label{maincoset}
g = e^{\im x^{\alpha\beta}P_{\alpha\beta} } e^{\im \theta^\alpha_i Q^i_\alpha}e^{\im \mpsi^\alpha_i(x,\theta) S^i_\alpha}.
\ee
Here, $x^{\alpha\beta}$ and $\theta^{\alpha}_i$ are the coordinates of the superspace, and $\mpsi^\alpha_i(x,\theta)$ are the Goldstone fermionic superfields. This is justified by their transformation laws. If the transformations in the coset space are induced by the left multiplication
\be\label{cosettrans}
g_0 g = g^\prime h, \;\;\; h = SO(1,5)\times SO(5),
\ee
the variations of $x$, $\theta$ and $\mpsi$ under unbroken and broken supersymmetry are
\bea\label{Qsusycosettrans}
g_Q = e^{\im \epsilon^{\alpha}_i Q^i_\alpha }&:& \;\; \delta_Q x^{\alpha\beta} = -\im \epsilon^{[\alpha}_i \theta^{\beta]i}, \; \delta_Q \theta^\alpha_i = \epsilon^\alpha_i, \; \delta_Q \mpsi^\alpha_i=0, \\
\label{Ssusycosettrans}
g_S = e^{\im \varepsilon^{\alpha}_i S^i_\alpha }&:& \;\; \delta_S x^{\alpha\beta} = -\im \epsilon^{[\alpha}_i \mpsi^{\beta]i}, \; \delta_S \mpsi^\alpha_i= \varepsilon^\alpha_i, \; \delta_S \theta^\alpha_i =0.
\eea
As expected, $x^{\alpha\beta}$ and $\theta^{\alpha}_i$ transform with respect to unbroken supersymmetry as the coordinates of the superspace, and the $\mpsi^\alpha_i$ remains inert. Conversely, $\theta^{\alpha}_i$ are not touched by broken supersymmetry, while the variations of $\mpsi^\alpha_i$ and $x^{\alpha\beta}$ remind the transformation laws of the Goldstone fermion proposed by Volkov and Akulov \cite{neugold} in four dimensions.

The Maurer-Cartan differential form $\Omega = g^{-1}dg$ is invariant with respect to the $Q$ and $S$ transformations:
\bea\label{MCform}
g^{-1}dg = \im \triangle x^{\alpha\beta}P_{\alpha\beta}- \im d\theta^\alpha_i \,Q^i_\alpha -\im d\mpsi^\alpha_i \,S^i_\alpha, \;\; \triangle x^{\alpha\beta} = dx^{\alpha\beta}- \im d\theta^{[\alpha}_i\, \theta^{\beta] i } - \im  d\mpsi^{[\alpha}_i\, \mpsi^{\beta] i }. \nn
\eea
Expanding the differential of the arbitrary invariant function in terms of the forms $\triangle x^{\alpha\beta}$ and $d\theta^\alpha_i$, one may construct derivatives covariant with respect to both supersymmetries:
\bea\label{covder}
\nabla_{\alpha\beta} = \big( E^{-1}\big)_{\alpha\beta}{}^{\mu\nu}\partial_{\mu\nu}, \;\;  E_{\alpha\beta}{}^{\mu\nu} = \delta^{[\mu}_\alpha \delta^{\nu]}_\beta  - \im \partial_{\alpha\beta}\mpsi^{[\mu}_i \, \mpsi^{\nu] i}, \nn \\
\nabla^i_\alpha = D^i_\alpha + \im \nabla^i_\alpha \mpsi^\rho_m \, \mpsi^{m\sigma}\partial_{\rho\sigma}, \;\; D^i_\alpha = \frac{\partial}{\partial\theta^\alpha_i} +\im \theta^{i\beta}\partial_{\alpha\beta}. \nn
\eea
As $\big\{ D^i_\alpha, D^j_\beta \big\}=2\im \epsilon^{ij}\partial_{\alpha\beta}$, their (anti)commutation relations are
\bea\label{covdercom}
&\big\{ \nabla^i_\alpha, \nabla^j_\beta  \big\} =2\im \epsilon^{ij} \nabla_{\alpha\beta} + 2\im \nabla^i_\alpha \mpsi^\rho_k \, \nabla^j_\beta \mpsi^{\sigma k}\nabla_{\rho\sigma},& \\
&\big[ \nabla_{\alpha\beta}, \nabla^i_\gamma \big] = 2\im \nabla_{\alpha\beta}\mpsi^\rho_m \, \nabla^i_\gamma \mpsi^{\sigma m}\nabla_{\rho\sigma}, \;\; \big[ \nabla_{\alpha\beta}, \nabla_{\mu\nu}  \big]=-2\im \nabla_{\alpha\beta}\mpsi^{\rho}_k \, \nabla_{\mu\nu}\mpsi^{\sigma k}\, \nabla_{\rho\sigma}.&   \nn
\eea

\section{The $N=(1,0)$, $d=6$ vector multiplet}
Let us briefly recall the properties of the $N=(1,0)$, $d=6$ vector multiplet. It was considered in the $SU(2)$ non-covariant approach in \cite{orig6dvect} and \cite{HST6dsusy}. The $SU(2)$ covariant formulation can be found in \cite{6danomcurrent}.  The latter is most useful when the formalism of nonlinear realizations is used. In this case, the usual $N=(1,0)$, $d=6$ vector multiplet is given by the spinorial superfield $\mpsi^\alpha_i$, subjected to the following irreducibility conditions
\be\label{N1d6vect}
D^i_\alpha \mpsi^{\alpha}_i =0,  D^i_\alpha \mpsi^{j\beta} +D^j_\alpha \mpsi^{i\beta}= \frac{1}{2}\delta_\alpha^\beta D^i_\gamma \mpsi^{j\gamma}.
\ee
One can check that these conditions imply that only the following components of the multiplet are independent:
\be\label{linvectcomps}
\psi^\alpha_i = \mpsi^\alpha_i |_{\theta \rightarrow 0}, \;\; F_\alpha{}^\beta = D^i_\alpha \mpsi^\beta_i|_{\theta \rightarrow 0}, \;\; B^{ij} = D^i_\alpha \mpsi^{j\alpha}|_{\theta \rightarrow 0}.
\ee
Acting on the $\mpsi^\alpha_i$ field by two spinorial derivatives, one finds that the result always reduces to the spacetime derivatives of $\mpsi^\alpha_i$.

It should be noted that as a consequence of the constraints \p{N1d6vect} the component $F_\alpha{}^\beta$ satisfies the differential identities known as the Bianchi identities. They indirectly imply that the antisymmetric tensor $F_\alpha{}^\beta= D^i_\alpha \mpsi^\beta_i|_{\theta \rightarrow 0}$ is the strength of some vector potential.

The first identity can be obtained by acting by two derivatives on the condition $D^k_\gamma \mpsi^{\gamma}_k =0$:
\be\label{linBI1}
D^i_\alpha D^j_\beta \big(D^k_\gamma \mpsi^{\gamma}_k \big) =0 \; \Rightarrow \; \partial_{\alpha\gamma}F_\beta{}^\gamma + \partial_{\beta\gamma}F_\alpha{}^\gamma =0.
\ee
The second one is a bit trickier. Analyzing the expression $\epsilon^{\alpha\mu\nu\lambda}D^i_\mu\, D^j_\nu\, D^k_\lambda \mpsi^{\beta}_k$, one can note that its part, symmetric in $\alpha$, $\beta$, is proportional to $\epsilon^{ij}$:
\be\label{linBI02}
\epsilon^{\alpha\mu\nu\lambda}D^i_\mu\, D^j_\nu\, D^k_\lambda \mpsi^{\beta}_k +  (\alpha\leftrightarrow \beta) =4\im \big( \partial^{\alpha\gamma}D^k_\gamma \mpsi^{\beta}_k + \partial^{\beta\gamma}D^k_\gamma \mpsi^{\alpha}_k \big)\epsilon^{ij}, \;\; \partial^{\alpha\beta} = \frac{1}{2}\epsilon^{\alpha\beta\mu\nu}\partial_{\mu\nu}.
\ee
Multiplying this by $\epsilon_{ij}$ and using the fact that $\epsilon^{\alpha\beta\mu\nu}\epsilon_{ij}D^i_\mu D^j_\nu = \frac{1}{2}\epsilon^{\alpha\beta\mu\nu}\epsilon_{ij}\big\{D^i_\mu, D^j_\nu \big\}$, one finds the second identity
\be\label{linBI2}
\partial^{\alpha\gamma}F_\gamma{}^\beta + \partial^{\beta\gamma}F_\gamma{}^\alpha=0.
\ee
In the $d=6$ vector notation, these two identities can be recognized as self-dual and anti self-dual parts of the identity $\partial_{[A}F_{BC]}=0$:
\bea\label{linBIvect}
\partial_{\alpha\gamma}F_\beta{}^\gamma + \partial_{\beta\gamma}F_\alpha{}^\gamma =\frac{1}{2} \big(\gamma^{ABC} \big)_{\alpha\beta} \partial_{[A}F_{BC]}, \nn \\
\partial^{\alpha\gamma}F_\gamma{}^\beta + \partial^{\beta\gamma}F_\gamma{}^\alpha =-\frac{1}{2} \big({\tilde\gamma}{}^{ABC} \big)^{\alpha\beta} \partial_{[A}F_{BC]}.
\eea

To construct the $N=(2,0)$, $d=6$ Born-Infeld action, it is required to find a proper covariant  generalization of these constraints, which would be compatible with additional spontaneously broken supersymmetry. As the construction of the actions of the $N=2$, $d=4$ and $N=4$, $d=3$ Born-Infeld theories shows, in the case of the vector multiplets it is not sufficient to formally covariantize the constraints with respect to the broken supersymmetry only by replacing the spinor derivatives with fully covariant ones \p{covder}. It is also required to choose the constraints which are covariant with respect to the automorphism group of the considered superalgebra.

Actually, the irreducibility conditions should be covariantized with respect to only the $SO(4)$ subgroup of the whole automorphism group $SO(5)$. Moreover, the $SU(2)$ part of the $SO(4)$ is realized by the linear transformations which rotate the indices $i,j$, and to preserve this symmetry, it would be sufficient to keep the balance of these indices. The transformations of the coset $SO(4)/SU(2)$ are realized on the variables $x^{\alpha\beta}$, $\theta^\alpha_i$, $\mpsi^\alpha_i$ as
\be\label{so4su2transf1}
g_R = e^{\im a_{ij}R^{ij}} \; \Rightarrow \; \delta x^{\alpha\beta} =0, \;\; \delta \theta^\alpha_i = a_i^k \mpsi^\alpha_k, \;\; \delta\mpsi^\alpha_i = a_i^k \theta^\alpha_k, \;\; \mbox{as} \; \delta \triangle x^{\alpha\beta}=0.
\ee
Now one can immediately derive variations of the differential forms $\triangle x^{\alpha\beta}$, $d\theta^\alpha_i$, $d\mpsi^\alpha_i$ with respect to these transformations and, finally, of the derivatives of $\mpsi^\alpha_i$:
\bea\label{so4su2transf2}
&\delta d\mpsi^\alpha_i = a_i^k d\theta^\alpha_k = \delta \triangle x^{\mu\nu}\, \nabla_{\mu\nu}\mpsi^\alpha_i + \delta d\theta^{\beta}_j \, \nabla_{\beta}^j\mpsi^\alpha_i + \triangle x^{\mu\nu} \delta \nabla_{\mu\nu}\mpsi^\alpha_i + d\theta^{\beta}_j \, \delta\nabla_{\beta}^j\mpsi^\alpha_i& \Rightarrow \nn \\
&\delta \nabla^i_\alpha \mpsi^\beta_j = a^i_j \delta^\beta_\alpha - a_m^k \nabla^i_\alpha \mpsi^\gamma_k \, \nabla^m_\gamma \mpsi^\beta_j.&
\eea
It can be noted that $\delta \nabla^i_\alpha \mpsi^\beta_j$ experiences a shift by the transformation parameter under these transformations, though it affects only its trace part over the Lorentz indices symmetrized with respect to $i,j$, $\nabla^{(i}_\alpha \mpsi^{j)\alpha} $. The first component of this combination is the auxiliary field of the multiplet.

Using the transformation laws \p{so4su2transf2}, one can establish the covariant generalization of the constraints \p{N1d6vect}. The simplest task is to generalize the constraint $D^i_\alpha \mpsi^\alpha_i=0$. One can observe that
\bea\label{traceauttr}
&\delta \nabla^i_\alpha\mpsi^\alpha_i = -a_m^k \nabla^m_\gamma \mpsi^\alpha_i \,\nabla^i_\alpha \mpsi^\gamma_k \, \equiv - a_m^k \big( \nabla\mpsi^2 \big)_{\gamma k}{}^{m \gamma}, & \\
&\delta \big( \nabla\mpsi^3 \big)_{\gamma k}{}^{k \gamma} = 3 a_m^k \big( \nabla\mpsi^2 \big)_{\gamma k}{}^{m \gamma} -3 a_m^k \big( \nabla\mpsi^4 \big)_{\gamma k}{}^{m \gamma}, \; \mbox{e.t.c.}&\nn
\eea
Therefore, in the following matrix power series variations of each term mutually cancel each other:
\be\label{1condgen0}
\delta\Big(\big(\nabla\mpsi  + \frac{1}{3} (\nabla\mpsi)^3 + \frac{1}{5}(\nabla\mpsi)^5 + \ldots \big)_{\gamma m}{}^{m \gamma} \Big) = \delta \tr\big[\mbox{arctanh}\big(\nabla^i_\alpha \mpsi^\beta_j \big)\big] =0.
\ee
As $\tr\big[\mbox{arctanh}\big(\nabla^i_\alpha \mpsi^\beta_j \big)\big]$ reduces to $D^i_\alpha \mpsi^\alpha_i$ when all nonlinear terms are neglected, the condition
\be\label{1condgen}
\tr\big[\mbox{arctanh}\big(\nabla^i_\alpha \mpsi^\beta_j \big)\big] =0
\ee
is the suitable one.

The second irreducibility condition should be generalized as
\be\label{2condgenans}
\nabla^{(i}_\alpha \mpsi^{j)\beta} = \frac{1}{4}\mY_\alpha{}^\beta \nabla^{(i}_\gamma \mpsi^{j)\gamma}, \;\; \tr\mY=4, \; \mY_\alpha{}^\beta = \delta_\alpha^\beta+\ldots.\nn
\ee
Here the matrix $\mY_\alpha{}^\beta$ should depend on $\mV_\alpha{}^\beta = \nabla^i_\alpha\mpsi^\beta_i$ and $\mB^2=\mB^{ij}\mB_{ij}$, $\mB^{ij} = \nabla^{(i}_\gamma \mpsi^{j)\gamma}$. Their transformation laws could be readily extracted from \p{so4su2transf2}:
\be\label{VBtransR}
\delta_R \mV_\alpha{}^\beta = 2\big( a\cdot \mB\big) \mV_\alpha{}^\lambda \mY_\lambda{}^\beta, \;\; \delta_R \mB^{ij} =  a^{ij}\Big( 4 - \frac{1}{4}\tr\big( \mV^2 \big) - \frac{1}{32} \tr\big( \mY^2 \big) \mB^2  \Big) + \frac{1}{16}\tr\big( \mY^2 \big)\big( a\cdot \mB\big)\,\mB^{ij}.
\ee
Then collecting the coefficients of $a^{ij}$, $\mB^{ij}$ in the variation of \p{2condgenans}, one can find that
\bea\label{2condgeneq}
a^{ij}: && \delta_\alpha^\beta - \frac{1}{4}\big(\mV^2\big)_\alpha{}^\beta - \frac{1}{32}\mB^2 \big(\mY^2\big)_\alpha{}^\beta = \mY_\alpha{}^\beta \Big( 1 - \frac{1}{16}\tr (\mV^2) -\frac{1}{128} \tr\big(\mY^2\big)\mB^2 \Big), \nn\\
\mB^{ij}: && \frac{1}{16}\big( a\cdot \mB\big) \mY_\alpha{}^\lambda \mY_\lambda{}^\beta =  \delta_R \mY_\alpha{}^\beta + \frac{1}{64}\big( a\cdot \mB\big)\mY_\alpha{}^\beta \tr\big( \mY^2 \big).
\eea
As we want to find the on-shell identity for the field strength, it is sufficient to know the irreducibility conditions in the first approximation in $\mB^{ij}$, or $\mY_\alpha{}^\beta$ in the limit $\mB\rightarrow 0$. Then the second relation could be neglected, while the first one implies that
\be\label{Y}
\mY_\alpha{}^\beta  \approx \frac{\delta_\alpha{}^\beta -  \frac{1}{4} \big(\mV^2\big)_\alpha{}^\beta}{4 - \frac{1}{4}\tr\big( \mV^2 \big)}.
\ee
It is convenient to write the approximate irreducibility condition as
\be\label{2condgen}
\nabla^{(i}_\alpha \mpsi^{j)\beta} \approx \frac{\mZ_\alpha{}^\beta}{\tr \mZ} \nabla^{(i}_\gamma \mpsi^{j)\gamma}, \;\; \mZ_\alpha{}^\beta = \delta_\alpha{}^\beta - \frac{1}{4}\big(\mV^2\big)_\alpha{}^\beta.\nn
\ee
As these conditions are known only approximately, it is not possible to fully check their consistency. However, they are, at least partially, justified by the latter construction.

It should be noted that one can establish the covariance of the constraints with respect to $R^{ij}$ and $T^{ij}$ transformations but not others. For example, for any generator that mixes $Q$ and $S$, like the generator $R_0$, the transformation law for $\nabla^i_\alpha \mpsi^\beta_j$ will contain a shift by the transformation parameter. However, the irreducibility condition can be written as a relation that expresses the general superfield $\nabla^i_\alpha \mpsi^\beta_j$ in terms of the superfields $\mB^{ij}=\nabla^{(i}_\alpha \mpsi^{j)\beta} $, $\hat\mV_\alpha{}^\beta =\nabla^i_\alpha \mpsi^\beta_i - \frac{1}{4}\delta_\alpha^\beta \nabla^i_\gamma \mpsi^\gamma_i $, the first components of which are independent components of the multiplet:
\be\label{genirr}
\nabla^i_\alpha \mpsi^{j\beta} = G_\alpha^{ij\beta}\big(\mB^{km}, \hat\mV_\mu{}^\nu  \big).
\ee
As the variation of the left-hand side contains the shift term, the variation of the right-hand side should contain such a term, too. Therefore, it is possible to covariantize the identity only with respect to the generators which can be associated with the auxiliary field of the multiplet.

Let us also note that the first irreducibility condition \p{1condgen} remains nonlinear even in the on-shell limit $\mB^{ij} = \nabla^{(i}_\alpha \mpsi^{j)\alpha}=0$:
\be\label{1condgenonsh}
\tr\Big[\mbox{arctanh}\Big(\frac{1}{2}\mV_\alpha{}^\beta \Big) \Big] =0.
\ee
Therefore, $\tr\big( \mV \big)$ is not equal to zero, unlike the linear case. It remains a nontrivial component, though it is expressed in terms of other components. Interestingly, this condition can be reduced to a much simpler cubic equation with the use of the formula $\det e^A = e^{\tr A}$:
\bea\label{1condgenonsh2}
\tr\Big[\mbox{arctanh}\Big(\frac{1}{2}\mV_\alpha{}^\beta \Big) \Big]=0 \; \Rightarrow \; \det\Big( \frac{1+\frac{1}{2}\mV}{1 - \frac{1}{2}\mV}  \Big) =1 \; \Rightarrow \nn \\
 24 \tr\big(\mV\big)+ \big(\tr\big(\mV\big)\big)^3 -3 \tr\big(\mV\big)\tr\big(\mV^2\big)+2\tr\big(\mV^3\big)=0.
\eea
Also, the derivative of this condition implies that
\be\label{1condgender}
d\tr\Big[\mbox{arctanh}\Big(\frac{1}{2}\mV_\alpha{}^\beta\Big) \Big] =0\; \Rightarrow \; d\mV_\alpha{}^\beta \, \big( \mZ^{-1} \big)_\beta{}^\alpha =0.
\ee

\section{Bianchi identities}
With the irreducibility conditions found, it is possible to derive differential identities that are satisfied by the components $V_\alpha{}^\beta = \mV_\alpha{}^\beta |_{\theta \rightarrow 0}$. The derivation of the identities can be made simpler if one needs only the identities in the bosonic limit and with the auxiliary field eliminated by its equation of motion in the final result. To perform this task, one needs to take the irreducibility conditions in the lowest nontrivial approximation in $\mB^{ij}$ and perform differentiation neglecting $\mB^{ij}$ in all cases when less than two spinorial derivatives act on it. Much like the identities in the linear case, the first identity can be found by acting by two derivatives on one of the irreducibility conditions:
\be\label{BI1}
\nabla^i_\alpha \nabla^j_\beta \Big(  \tr\big[\mbox{arctanh}\big(\nabla^k_\mu \mpsi^\nu_m \big)\big]  \Big) = 0 \; \Rightarrow \; \big(BI\big)_{\alpha\beta} =  \Big(\partial_{\alpha\rho} V_\beta{}^\sigma +\frac{1}{4} V_\alpha{}^\mu V_\rho{}^\nu \partial_{\mu\nu} V_\beta{}^\sigma \Big) \big(Z^{-1}\big)_\sigma{}^\rho + (\alpha\leftrightarrow \beta) =0.
\ee
The second identity can be found by the analysis of the expression $\epsilon^{\alpha\mu\nu\lambda}\nabla^i_\mu \nabla^j_\nu \nabla^k_\lambda \mpsi^\beta_k$:
\be\label{BItilde1}
\big( {\widetilde{BI}} \big)^{\alpha\beta} =\epsilon^{\alpha\mu\nu\lambda}\Big(\partial_{\mu\nu}V_\lambda{}^\gamma +\frac{1}{4}V_\mu{}^\rho V_\nu{}^\sigma \partial_{\rho\sigma} V_\lambda{}^\gamma\Big)\big(Z^{-1}\big)_\gamma{}^\beta + (\alpha\leftrightarrow \beta) =0.
\ee

These identities should be equivalent to the usual ones. This requires that, in particular, the matrices $M_{(\alpha\beta)}{}^{(\mu\nu)}$, $N_{(\alpha\beta)(\mu\nu)}$, ${\widetilde M}_{(\mu\nu)}{}^{(\alpha\beta)}$, ${\widetilde N}^{(\alpha\beta)(\mu\nu)}$ should exist, such that
\bea\label{combBIspinor}
\partial_{\alpha\gamma}F_\beta{}^\gamma + \partial_{\beta\gamma}F_\alpha{}^\gamma &=& M_{(\alpha\beta)}{}^{(\mu\nu)} \big(BI\big)_{\mu\nu} +  N_{(\alpha\beta)(\mu\nu)}\big( {\widetilde{BI}} \big)^{\mu\nu}, \nn \\
\partial^{\alpha\gamma}F_\gamma{}^\beta + \partial^{\beta\gamma}F_\gamma{}^\alpha &=& {\widetilde M}_{(\mu\nu)}{}^{(\alpha\beta)} \big( {\widetilde{BI}} \big){}^{\mu\nu} +  {\widetilde N}^{(\alpha\beta)(\mu\nu)}\big(BI\big)_{\mu\nu}.
\eea
In principle, one may treat $F_\alpha{}^\beta$ as a polynomial of degree $3$ in $V_\alpha{}^\beta$, the matrices $M_{(\alpha\beta)}{}^{(\mu\nu)}$, $N_{(\alpha\beta)(\mu\nu)}$ - as double polynomials, and equate both sides of relations \p{combBIspinor}. This approach, however, is very tedious and does not shed light on the nature of the matrices $M$, $N$. Additionally, it requires to analyze two separate identities.

To avoid these difficulties, one should rewrite the identities in the vector notation. To additionally simplify these relations, one may note that in both of them the derivatives are found as part of the combination $\mD_{\alpha\beta} = \partial_{\alpha\beta} + \frac{1}{4}V_\alpha{}^\mu V_\beta{}^\nu\partial_{\mu\nu} \equiv  \Sigma  _{\alpha\beta}{}^{\mu\nu}\partial_{\mu\nu}$. Then one can represent $V_\alpha{}^\beta$ and $\big( Z^{-1}  \big)_{\alpha}{}^{\beta}$ as
\bea\label{VZinvvect}
&V_\alpha{}^\beta = A \delta_\alpha^\beta +\frac{1}{2}\big( \gamma^{AB}  \big)_\alpha{}^\beta V_{AB}, \;\; \big( Z^{-1}  \big)_{\alpha}{}^{\beta} = G_0 \big( \delta_\alpha^\beta +\frac{1}{2}\big( \gamma^{AB}  \big)_\alpha{}^\beta G_{AB}  \big), \mbox{where}&\nn \\
&G_{AB} = -\frac{1}{2}\frac{A V_{AB} +\frac{1}{8}\epsilon_{ABCDMN}V^{CD}V^{MN}}{1 + \frac{3}{4}A^2 + \frac{1}{8}V_{CD}V^{CD}}& \eea
and $G_0$ could be canceled from the identities. The relation on the components of $V_\alpha{}^\beta$ \p{1condgenonsh2} now implies
\be\label{1condgenonsh3}
\epsilon^{ABCDMN}V_{AB}V_{CD}V_{MN} +96 A + 24 A^3 + 12 A V_{CD}V^{CD} =0.
\ee

With the help of \p{VZinvvect}, two identities \p{BI1}, \p{BItilde1} can be written as follows:
\bea\label{BI1tilde1vect}
\big( \gamma^{ABC}  \big)_{\alpha\beta} \big( \mD_A V_{BC} + G_{BC} \mD_A A -2 \mD_A V_{BK} G_C{}^K - \mD^K V_{KA}\,G_{BC} - \mD^K V_{AB}\,G_{KC} \big) =0, \nn \\
\big( {\tilde\gamma}^{ABC}  \big)^{\alpha\beta} \big( \mD_A V_{BC} + G_{BC} \mD_A A -2 \mD_A V_{BK} G_C{}^K +\mD^K V_{KA}\,G_{BC}+ \mD^K V_{AB}\, G_{KC} \big) =0.
\eea
Here $\mD_A = -\frac{1}{2}\big({\tilde\gamma}^A \big)^{\alpha\beta} \mD_{\alpha\beta} \equiv \Sigma_A{}^B \partial_B$,
\be\label{Sigmamatr}
\Sigma_A{}^B = \big( 1+\frac{1}{4} A^2  + \frac{1}{8}V_{KL}V^{KL}  \big) \delta_A{}^B + \frac{1}{2}A V_A{}^B + \frac{1}{2}V_{AK}V^{KB} + \frac{1}{16}\epsilon_A{}^{BCDMN}V_{CD}V_{MN}.
\ee

Taking into account the self-duality properties of $\big( \gamma^{ABC}  \big)_{\alpha\beta}$, $\big( {\tilde\gamma}^{ABC}  \big)^{\alpha\beta}$ (see Appendix), two relations \p{BI1tilde1vect} are equivalent to the single one
\be\label{BIvect}
\mD_{[A} V_{BC]} +  \mD_{[A} A \, G_{BC]} -2 \mD_{[A} V_{B}{}^K G_{C]K} - \frac{1}{6} \epsilon_{ABCMNP}\big( \mD_K V^{KM}G^{NP} + \mD_K V^{MN}G^{KP} \big) =0.
\ee
Using the identity $\epsilon_{[ABCMNP}\mD_{K]}=0$, \p{BIvect} can also be presented as
\be\label{BIvect2}
\mD_{[A} V_{BC]} +  \mD_{[A} A \, G_{BC]} -2 \mD_{[A} V_{B}{}^K G_{C]K} - \frac{1}{4} \epsilon_{KMNP[BC} \mD_{A]} V^{KM} G^{NP} =0.
\ee
It is now clear that this identity should be multiplied by three matrices $\big(  \Sigma^{-1}\big)_{A^\prime}{}^A \big(  \Sigma^{-1}\big)_{B^\prime}{}^B \big(  \Sigma^{-1}\big)_{C^\prime}{}^C $ to be brought to the standard form because it is one and only way to make the indices of all derivatives $\partial_{A^\prime}$ free, as in the canonical identity. To prove exactly that after this multiplication \p{BIvect2} finally acquires the expected form, it is convenient to introduce the matrix
\bea\label{Phimatrix}
\Phi_{AB} = \big( 1 - \frac{1}{4} A^2 + \frac{1}{8}V_{KL}V^{KL}  \big)V_{AB} + \frac{1}{2}V_{AC}V^{CD}V_{DB}- \frac{A}{16}\epsilon_{AB}{}^{CDMN}V_{CD}V_{MN}, \nn \\
  \Phi_{AC}\big( \Sigma^{-1} \big)_B{}^C = A \eta_{AB} + V_{AB}.
\eea
In terms of this matrix, identity \p{BIvect2} reads
\bea\label{BIvectPhi}
&&\mD_{[A}\Phi_{BC]} + A^2 \mD_{[A}V_{BC]} + \frac{1}{4}A\epsilon_{[BC}{}^{MNPQ}\mD_{A]}V_{MN}\, V_{PQ} + A \mD_{[A}V_B{}^K \, V_{C]K}- \frac{1}{4}V_{[BC}\mD_{A]}\big( V_{KL}V^{KL}  \big)+\nn \\
&&+ \frac{1}{4}\epsilon_{MNPQK[C}\mD_A V^{MN}\, V^{PQ}V_{B]}{}^K - \mD_{[A}V_B{}^K V_{C]L}V^L{}_K - V_{K[B}\mD_A V^{KL} \, V_{C]L}=0.
\eea
After the multiplication by $\big(  \Sigma^{-1}\big)_{A^\prime}{}^A \big(  \Sigma^{-1}\big)_{B^\prime}{}^B \big(  \Sigma^{-1}\big)_{C^\prime}{}^C $ and the integration by parts, the first term can be presented as
\be\label{BIvectPhifirstterm}
\partial_{[A^\prime} \big( \big( \Sigma^{-1} \big)_{B^\prime}{}^B \big( \Sigma^{-1} \big)_{C^\prime]}{}^C \Phi_{BC} \big) + 2 \big(  \Sigma^{-1}\big)_{[A^\prime}{}^A \big(  \Sigma^{-1}\big)_{B^\prime}{}^B \big(  \Sigma^{-1}\big)_{C^\prime]}{}^C \big( \Phi_{KC}\big(  \Sigma^{-1}\big)_{L}{}^K  \mD_A \Sigma_B{}^L   \big).
\ee
Using the properties of $\Phi_{AB}$ \p{Phimatrix} and explicitly taking the derivative of $\Sigma_B{}^L$, one may find that the generated terms cancel all the extra terms in identity \p{BIvectPhi}. Therefore, the right identity reads $\partial_{[A}F_{BC]}=0$, where
\bea\label{physbosFvect}
F_{AB} &=& \big( \Sigma^{-1} \big)_{A}{}^C \big( \Sigma^{-1} \big)_{B}{}^D\Phi_{CD}= \big( \Sigma^{-1} \big)_{A}{}^C \big(   A \eta_{CB} + V_{CB}\big)= \nn \\
&=&\frac{\big(  1+\frac{1}{4}A^2 -\frac{1}{8}V_{CD}V^{CD} \big)V_{AB} + \frac{1}{16}A \epsilon_{ABCDPQ}V^{CD}V^{PQ} - \frac{1}{2}V_{AC}V^{CD}V_{DB}}{1 +A^2 + \frac{3}{16}A^4 + \frac{1}{16}A^2 V_{KL}V^{KL} - \frac{1}{16}V_{KL}V^{LM}V_{MN}V^{NK}+ \frac{1}{64}\big(  V_{KL}V^{KL} \big)^2}.
\eea

For further considerations, it is useful to write it down in the spinor notation:
\be\label{physbosF}
F_\alpha{}^\beta = \frac{\frac{1}{2}\tr\big(V\big)\delta_\alpha^\beta + \big( 1 + \frac{1}{8}\big(\tr\big(V\big)\big)^2 - \frac{1}{8}\tr\big(V^2\big) \big)V_\alpha^{\beta} - \frac{1}{4}\tr\big(V\big) \big(V^2\big)_\alpha{}^\beta +\frac{1}{4}\big(V^3\big)_{\alpha}{}^\beta}{1 + \frac{1}{4}\big(\tr\big(V\big)\big)^2 +\frac{1}{128}\big(\tr\big(V\big)\big)^4 - \frac{1}{128}\big(\tr\big(V^2\big)\big)^2 - \frac{1}{64}\big(\tr\big(V\big)\big)^2 \tr\big(V^2\big) + \frac{1}{64} \tr\big(V^4\big)}.
\ee
Here  relation \p{1condgenonsh2} was used to express $\tr\big(V^3\big)$ in terms of $\tr\big(V \big)$ and $\tr\big( V^2 \big)$. Let us also note that the numerator of \p{physbosF} can be written as
\be\label{physbosFnum}
\sqrt{\det Z} \Big( \big( Z^{-1}\big)_\alpha{}^\lambda V_\lambda{}^\beta - \frac{1}{4} \delta_\alpha^\beta \big( Z^{-1}\big)_\rho{}^\sigma V_\sigma{}^\rho  \Big).
\ee

\section{Broken supersymmetry}
The component approach to the actions with broken supersymmetry involves the construction of the ansatz for the action  invariant with respect to broken supersymmetry by modifying the measure and the derivatives in the bosonic action and adding the Wess-Zumino terms and checking its invariance with respect to unbroken supersymmetry. As $\theta^{\alpha}_i$ are invariant with respect to broken supersymmetry, the necessary transformation laws and invariant forms can be obtained from \p{Ssusycosettrans}, \p{MCform} in the limit $\theta \rightarrow 0$.  Therefore, the covariant derivative which acts on the components reads
\be\label{cEcD}
\cD_{\alpha\beta} = \big( \cE^{-1}\big)_{\alpha\beta}{}^{\mu\nu}\partial_{\mu\nu}, \;\; \cE_{\alpha\beta}{}^{\mu\nu} = E_{\alpha\beta}{}^{\mu\nu}|_{\theta \rightarrow 0} = \delta_\alpha^{[\mu} \delta_\beta^{\nu]} - \im \partial_{\alpha\beta} \psi^{[\mu}_i \, \psi^{\nu] i}.
\ee
It is also useful to rewrite the derivatives and the matrices in the vector notation
\be\label{cEcD2}
\cD_A = \big( \cE^{-1}  \big)_A{}^B \partial_B, \;\; \cE_A{}^B = \delta_A^B - \frac{\im}{2} \partial_A \psi^\rho_i \, \psi^{\sigma i} \big( \gamma^B \big)_{\rho\sigma}.
\ee
The active transformation laws of the fields and the vielbein in the vector notation are
\bea\label{Sactvar}
&\delta^\star_S\psi^\alpha_i = \varepsilon^\alpha_i + U^A \partial_A \psi^\alpha_i, \;\; \delta^\star_S F_{AB} = U^C\partial_C F_{AB}, \;\; U^A = \frac{\im}{2}\varepsilon^\mu_i \psi^{\nu i}\big( \gamma^A \big)_{\mu\nu},& \nn \\
&\delta^\star_S \cE_A{}^B = \partial_A U^C \cE_C{}^B + U^C\partial_C \cE_A{}^B, \;\; \delta_S^\star \det\cE = \partial_A \big( U^A \det\cE\big), \;\; \delta_S^\star \cD_A \psi^\alpha_i = U^C \partial_C  \cD_A \psi^\alpha_i.&
\eea
The invariant measure is, therefore, $d^6 x \det\cE$.

With these transformations at hand, one can restore the fermionic contributions to the field strength. As is easy to note, the unmodified Bianchi identity $\partial_{[A}F_{BC]}=0$ is not invariant (due to variation of $F_{AB}$, when active transformations are considered, or due to nontrivial variation of $x^A$ if one considers usual transformations). Therefore, the true physical field strength ${\cal F}_{AB}$ should have another transformation law with respect to broken supersymmetry. The comparison with the $N=2$, $d=4$ Born-Infeld theory suggests that the right field strength is ${\cal F}_{AB} = \cE_A{}^C \cE_B{}^D F_{CD}$. Varying the expression $\partial_{[A}{\cal F}_{BC]}$, one may find that it transforms proportionally to itself and its derivatives:
\bea\label{trueBIvar}
\delta^\star_S \partial_{[A} {\cal F}_{BC]} = -2 \partial_{[B}U^K\partial_A {\cal F}_{C]K}+\partial_K {\cal F}_{[BC} \partial_{A]}U^K + U^K\partial_K \big(\partial_{[A}{\cal F}_{BC]}\big) = \nn \\
=\partial_{[B}U^K \partial_{A}{\cal F}_{KC]} + \partial_K U^K \, \partial_{[A} {\cal F}_{BC]} + U^K\partial_K \big(\partial_{[A}{\cal F}_{BC]}\big).
\eea
Therefore, the identity $\partial_{[A}{\cal F}_{BC]}=0$ is compatible with broken supersymmetry in all approximations in the fermions.

The simple transformation law of $F_{AB}$, in comparison with ${\cal F}_{AB}$, suggests that the bosonic core of the action should be generalized as
\be\label{S0}
S_0 = -\int d^6 x \det \cE \Big( C_1 + \sqrt{-\det\big(\eta_{AB}+F_{AB}\big)} \Big).
\ee
Comparing the lowest nontrivial limit of \p{S0} with the free action, one may immediately determine $C_1 =1$. It can be rewritten in terms of the variables $V_\alpha{}^\beta$ too,
\be\label{S01}
S_0 = -\int d^6 x \frac{ 2\det\cE \big( 1+ \frac{1}{16}\big(\tr\big(V\big)\big)^2 - \frac{1}{16}\tr\big(V^2\big) \big)}{1 + \frac{1}{4}\big(\tr\big(V\big)\big)^2 +\frac{1}{128}\big(\tr\big(V\big)\big)^4 - \frac{1}{128}\big(\tr\big(V^2\big)\big)^2 - \frac{1}{64}\big(\tr\big(V\big)\big)^2 \tr\big(V^2\big) + \frac{1}{64} \tr\big(V^4\big) }.
\ee

The Wess-Zumino term also should be constructed. As the main action \p{S01} involves the terms of even power in the fermions and in the bosons, the Wess-Zumino term which could make a useful contribution to the action should also  be quadratic in the field strengths and at least quadratic in the fermions. Also, its variation with respect to broken supersymmetry transformations \p{Sactvar} should reduce to the full derivative. Therefore, one can expect this term to approximately read
\be\label{LWZappr}
{\cal L}_{WZ} \approx \im\epsilon^{ABCDMN}\psi^\alpha_i \, \partial_A \psi^{\beta i}  \big(\gamma_B \big)_{\alpha\beta}F_{CD}F_{MN}.\nn
\ee
Indeed, in the lowest order in the fermions, the only field which transforms is $\psi^\alpha_i$ without the derivative, $\delta_S \psi^\alpha_i \sim \varepsilon^\alpha_i$. Then $\delta_S {\cal L}_{WZ}$ can be integrated by parts, and the appearing terms with the derivatives of $F_{AB}$ will vanish due to the Bianchi identity.

By adding more terms with the fermions, one can make the Wess-Zumino action invariant with respect to broken supersymmetry in all approximations in the fermions:
\be\label{LWZ}
{\cal L}_{WZ} =\im \det\cE \epsilon^{ABCDMN}\psi^\alpha_i \, \cD_A \psi^{\beta i}  \big(\cE^{-1}\big)_B{}^K\big(\gamma_K \big)_{\alpha\beta}F_{CD}F_{MN}.
\ee
Indeed, varying this term with respect to transformations \p{Sactvar}, one can find that
\bea\label{LWZvar}
\delta^\star_S {\cal L}_{WZ} &=&\im \det\cE \epsilon^{ABCDMN}\varepsilon^\alpha_i \, \cD_A \psi^{\beta i}\big(\cE^{-1}\big)_B{}^K\big(\gamma_K \big)_{\alpha\beta}F_{CD}F_{MN} - \nn \\&& - \im \det\cE \epsilon^{ABCDMN}\psi^\alpha_i \, \cD_A \psi^{\beta i}\cD_B{}U^K\big(\gamma_K \big)_{\alpha\beta}F_{CD}F_{MN}  + \partial_A \big( U^A {\cal L}_{WZ} \big) = \nn \\
&=&  \im \epsilon^{ABCDMN}\varepsilon^\alpha_i \, \partial_A \psi^{\beta i}\big(\gamma_B \big)_{\alpha\beta} \cE_C{}^K \cE_D{}^L F_{KL}\, \cE_M{}^P \cE_N{}^Q F_{PQ}+ \partial_A \big( U^A {\cal L}_{WZ} \big) -\\&&- \epsilon^{ABCDMN}\psi^\alpha_i \, \partial_A \psi^{\beta i}\varepsilon^\mu_j \partial_B \psi^{\nu j}\epsilon_{\alpha\beta\mu\nu}\cE_C{}^K \cE_D{}^L F_{KL}\, \cE_M{}^P \cE_N{}^Q F_{PQ}  =\nn \\
&=& \im\epsilon^{ABCDMN}\varepsilon^\alpha_i \, \partial_A \psi^{\beta i}\big(\gamma_B \big)_{\alpha\beta} {\cal F}_{CD}\, {\cal F}_{MN}+  \partial_A \big( U^A {\cal L}_{WZ} \big)-\nn \\&&
- \frac{1}{3} \epsilon^{ABCDMN}\epsilon_{\alpha\beta\mu\nu}\partial_B \big(\varepsilon^\mu_j \psi^{\nu j} \,  \psi^{\alpha }_i \partial_A\psi^{\beta i} \big) {\cal F}_{CD}\, {\cal F}_{MN},\nn
\eea
which is full divergence due to the previously established Bianchi identities. In the last line we used two relations
\bea\label{LWZvaraux}
\psi^\alpha_i \partial_A \psi^{\beta i} \varepsilon^\mu_j \partial_B \psi^{\nu j}\epsilon^{ABCDMN}\epsilon_{\alpha\beta\mu\nu} = \epsilon^{ABCDMN}\epsilon_{\alpha\beta\mu\nu}  \partial_B\big(  \psi^\alpha_i \partial_A \psi^{\beta i} \varepsilon^\mu_j  \psi^{\nu j}\big)  + \nn \\+ \epsilon^{ABCDMN}\epsilon_{\alpha\beta\mu\nu}  \partial_A  \psi^\alpha_i \partial_B \psi^{\beta i} \varepsilon^\mu_j  \psi^{\nu j} \mbox{ and} \\
\psi^\alpha_i \partial_A \psi^{\beta i} \varepsilon^\mu_j \partial_B \psi^{\nu j}\epsilon^{ABCDMN}\epsilon_{\alpha\beta\mu\nu} = - \frac{1}{2}\varepsilon_j^{\mu}\psi^{\alpha j} \partial_A \psi^{\beta }_i \partial_B \psi^{\nu i}\epsilon^{ABCDMN}\epsilon_{\alpha\beta\mu\nu}.\nn
\eea

It would be useful to rewrite the Wess-Zumino term in the spinor notation. In the lowest approximation in the fermions it reads
\be\label{LWZspin}
{\cal L}_{WZ} \approx  4 \im \psi^{\alpha}_i \partial_{\lambda\beta}\psi^{\beta i} \big( F^2  \big)_\alpha{}^\lambda  +4\im  \psi^{\alpha}_i \partial_{\alpha\lambda}\psi^{\beta i} \big( F^2  \big)_\beta{}^\lambda  - 2\im \tr\big( F^2  \big) \psi^{\alpha}_i \partial_{\alpha\beta}\psi^{\beta i} .
\ee

\section{Unbroken supersymmetry}
The last point in constructing the action is checking its invariance with respect to unbroken supersymmetry. As one of the coefficients in the action \p{S0} was already fixed by the invariance with respect to unbroken supersymmetry in lowest approximation, only one free constant $C_{WZ}$ remains:
\be\label{actans}
S = S_0 + C_{WZ}\,S_{WZ}.
\ee
It should be determined by the invariance with respect to the complete unbroken supersymmetry transformations taken in the lowest approximation in $\psi^\alpha_i$. The transformations of the components can be derived with the help of the formula
\be\label{Qtrans}
\delta^\star_Q f = -\epsilon^\alpha_i D^i_\alpha \mf|_{\theta\rightarrow 0} \equiv - \epsilon^\alpha_i \nabla^i_\alpha \mf|_{\theta\rightarrow 0} + H^{\mu\nu}\partial_{\mu\nu} f, \; H^{\mu\nu} = \frac{\im}{2}\epsilon^\lambda_i V_{\lambda}{}^{[\mu}\psi^{\nu] i}.
\ee
As we plan to prove the invariance of the action in the first order in the fermions, the $H^{\mu\nu}$ terms are not relevant and all broken supersymmetry covariant derivatives can be replaced with the usual ones $\cD_{\alpha\beta}\rightarrow \partial_{\alpha\beta}$. The transformations of the basic components in the lowest approximation in the fermions then read
\bea\label{Qtranscomp}
\delta_Q^\star \psi^\alpha_i &\approx& -\frac{1}{2}\epsilon^{\beta i}V_\beta{}^\alpha , \\
\delta_Q^\star V_{\alpha}{}^\beta &\approx& -4\im \epsilon^\gamma_i \mD_{\gamma\alpha}\psi^{i\beta}-4\im \epsilon^\gamma_i Z_\gamma{}^\beta  \mD_{\alpha\mu}\psi^{\nu i}  \big( Z^{-1} \big)_\nu{}^\mu + 2\im Z_\alpha{}^\beta \epsilon^\gamma_i \mD_{\gamma\mu}\psi^{\nu i}\big( Z^{-1} \big)_\nu{}^\mu.\nn
\eea
The variation of $\det\cE$ is relatively simple,
\be\label{deltaQdetCE}
\delta^\star_Q \det\cE \approx -\im \epsilon^\lambda_i V_\lambda{}^\mu \partial_{\mu\nu}\psi^{\nu i},
\ee
while the variations of the basic bosonic invariants whose action depends on, $\tr\big(V\big)$, $\tr\big(V^2\big)$ and $\tr\big(V^4\big)$, are too large to be written explicitly.  Also, only the fermions have to be varied in the Wess-Zumino term \p{LWZ}, where, up to the full derivative,
\bea\label{deltaQLWZ}
\delta^\star_Q {\cal L}_{WZ} &\approx& 2 \im \epsilon^{ABCDMN}\delta^\star_Q \psi^\alpha_i \, \partial_A \psi^{\beta i}\big(\gamma_B \big)_{\alpha\beta}F_{CD}F_{MN} = \nn \\
&=&  8 \im \delta^\star_Q\psi^{\alpha}_i \partial_{\lambda\beta}\psi^{\beta i} \big( F^2  \big)_\alpha{}^\lambda  +8\im  \delta^\star_Q\psi^{\alpha}_i \partial_{\alpha\lambda}\psi^{\beta i} \big( F^2  \big)_\beta{}^\lambda  - 4\im \tr\big( F^2  \big) \delta^\star_Q\psi^{\alpha}_i \partial_{\alpha\beta}\psi^{\beta i}.
\eea

Then the whole variation of the action \p{actans} can be written as a sum of terms with the structure $\Phi_{(k)(m)} = \epsilon^\alpha_i \big( V^k \big)_{\alpha}{}^\rho \partial_{\rho\sigma}\psi^{i\beta} \big( V^m \big)_{\beta}{}^\sigma$, where $k,m=0,\ldots,3$, with scalar coefficients. They can be rewritten in terms of ${\widetilde\Phi}_{(k)(m)} = \epsilon^\alpha_i \big( F^k \big)_{\alpha}{}^\rho \partial_{\rho\sigma}\psi^{i\beta} \big( F^m \big)_{\beta}{}^\sigma$, as $\big(F^k\big)_{\alpha}{}^\beta$ can also be written as polynomials in $\big( V^m \big)_{\beta}{}^\sigma$ and, therefore, $\Phi_{(k)(m)}$ and ${\widetilde\Phi}_{(k)(m)}$ are {\it linearly} related to each other by the matrix, the elements of which are functions of traces of powers of $V_\alpha{}^\beta$. It can be noted that if $C_{WZ}=-\frac{1}{16}$, the variation of the Lagrangian in \p{actans} can be cast into a relatively simple form
\bea\label{lagrvar}
\delta^\star_Q {\cal L}&=& \im \Big( 1 + \frac{1}{8}\tr\big( F^2 \big)  \Big){\widetilde\Phi}_{(1)(0)}- \im \Big( 1 - \frac{1}{8}\tr\big( F^2 \big)  \Big){\widetilde\Phi}_{(0)(1)} - \nn \\&&- \frac{\im}{4}\Big( {\widetilde\Phi}_{(0)(3)} + {\widetilde\Phi}_{(3)(0)} +{\widetilde\Phi}_{(1)(2)}+ {\widetilde\Phi}_{(2)(1)} \Big) + \frac{\im}{12}\tr\big( F^3 \big){\widetilde\Phi}_{(0)(0)}.
\eea
Though it is far from obvious, the terms with the third power of $F_{\alpha}{}^\beta$ actually cancel out. To prove this, let us write all pieces of \p{lagrvar} in the vector notation:
\bea\label{termslagrvar}
{\widetilde\Phi}_{(1)(0)}-{\widetilde\Phi}_{(0)(1)} &=& -\frac{1}{2} \big( \gamma^{ABC} \big)_{\alpha\beta} \epsilon^{\alpha}_i \partial_C \psi^{i\beta}F_{AB}, \nn \\
\tr\big( F^3 \big){\widetilde\Phi}_{(0)(0)} &=& \frac{1}{4} \epsilon^{ABMNPQ}F_{AB}F_{MN}F_{PQ}\epsilon^\alpha_i \partial_C \psi^{\beta i} \big( \gamma^C \big)_{\alpha\beta}, \nn \\
\tr\big(F^2  \big) \big( {\widetilde\Phi}_{(1)(0)}+{\widetilde\Phi}_{(0)(1)}  \big)& = &2 F_{KL}F^{KL} \epsilon^\alpha_i \partial_A \psi^{\beta i} F^{AB}\big( \gamma_B \big)_{\alpha\beta}, \nn \\
{\widetilde\Phi}_{(3)(0)}+{\widetilde\Phi}_{(0)(3)} &=& 2\epsilon^\alpha_i \partial_A \psi^{\beta i} F^{AC}F_{CD}F^{DB}\big( \gamma_B \big)_{\alpha\beta} + \frac{3}{2}F_{KL}F^{KL} \epsilon^\alpha_i \partial_A \psi^{\beta i} F^{AB}\big( \gamma_B \big)_{\alpha\beta} +\nn \\
&&+ \frac{1}{8}\epsilon^{ABMNPQ}F_{AB}F_{MN}F_{PQ}\epsilon^\alpha_i \partial_C \psi^{\beta i} \big( \gamma^C \big)_{\alpha\beta},  \\
{\widetilde\Phi}_{(2)(1)}+{\widetilde\Phi}_{(1)(2)} &=& -2\epsilon^\alpha_i \partial_A \psi^{\beta i} F^{AC}F_{CD}F^{DB}\big( \gamma_B \big)_{\alpha\beta} -\frac{1}{2}F_{KL}F^{KL} \epsilon^\alpha_i \partial_A \psi^{\beta i} F^{AB}\big( \gamma_B \big)_{\alpha\beta} - \nn \\
&&-\frac{1}{24}\epsilon^{ABMNPQ}F_{AB}F_{MN}F_{PQ}\epsilon^\alpha_i \partial_C \psi^{\beta i} \big( \gamma^C \big)_{\alpha\beta}. \nn
\eea
It is now evident that all cubic terms cancel out, and the only linear term is  full divergence due to the bosonic Bianchi identity.

Therefore, the final action reads
\bea\label{Sfin}
S &=&-\int d^6 x \det \cE \big(1 + \sqrt{-\det\big( \eta_{AB} + F_{AB} \big)}  \big) - \nn \\
&&-\frac{\im}{16}\int d^6 x\det\cE \epsilon^{ABCDMN}\psi^\alpha_i \, \cD_A \psi^{\beta i}  \big(\cE^{-1}\big)_B{}^K\big(\gamma_K \big)_{\alpha\beta}F_{CD}F_{MN}.
\eea

\section{Reduction to four dimensional theory}
With the action of the $N=2$, $d=6$ theory at hand \p{Sfin}, it would be natural to find the action of the related $N=4$, $d=4$ Born-Infeld theory by the dimensional reduction \cite{ketov1}, \cite{nlselfdual}. It implies taking all the fields independent of two coordinates, $x^4$ and $x^5$. As the physical field strength can be represented as  antisymmetrized derivative of the potential, ${\cal F}_{AB} = \partial_A {\cal A}_B - \partial_B {\cal A}_A$, the aforementioned reduction also implies ${\cal F}_{45}=0$.

Two most important points should be studied here, the invariance of the reduced action with respect to both broken and unbroken $N=2$, $d=4$ supersymmetries, and the self-duality of the reduced action \cite{nlselfdual}. Throughout this section, we denote the four dimensional vector indices by $\uA,\uB =0,\ldots,3$ and the four dimensional spinor indices by $\ualpha,\ubeta=1,2$ and $\dot\ualpha,\dot\ubeta=\dot 1, \dot 2$.

One can note that only the law of reduction of the field strength $\cF_{45}=0$ is simple, while the reduction of $V_{AB}$ is considerably more complicated; it is not even known how to explicitly express $V_{AB}$ in terms of $F_{AB}$. As the unbroken supersymmetry transformation laws are written in terms of $V_{AB}$, it would be desirable to avoid explicit proof of invariance of the action and just verify that the reduction is compatible with both supersymmetries. This means that variations of all fields that vanish upon the reduction should also reduce to zero and the reduced transformations should form a closed superalgebra. To clarify this point, let us consider the general form of transformations:
\be\label{XYtr}
\delta X^{(i)} = \epsilon^{(\alpha)}K^{(i)}_{(\alpha)}(X,Y), \;\; \delta Y^{(a)} = \epsilon^{(\alpha)}L^{(a)}_{(\alpha)}(X,Y).
\ee
Here we split all the basic fields and their derivatives into two sets: $X^{(i)}$ fields, such as $\psi^{\alpha}_i$, $\partial_1 \psi^{\alpha}_i$, $\cF_{12}$, just lose their dependence on $x^4$, $x^5$ upon the reduction, while $Y^{(a)}$ fields, like $\cF_{45}$, $\partial_4 \psi^{\alpha}_i$, vanish upon the reduction. The parameter $\epsilon^{(\alpha)}$ combines the parameters of both broken and unbroken supersymmetries.

To keep the reduced Lagrangian supersymmetric, it is necessary to have $L \rightarrow 0$ if $Y\rightarrow 0$. Indeed, the variation of the six dimensional Lagrangian, which depends on $X,Y$ and the variation of the four dimensional reduced Lagrangian, which depends on $X$ only, read
\bea\label{Lagrvar}
\delta \cL &=& \epsilon^{(\alpha)}K^{(i)}_{(\alpha)}(X,Y)\frac{\partial \cL}{\partial X^{(i)}} + \epsilon^{(\alpha)}L^{(a)}_{(\alpha)}(X,Y)\frac{\partial \cL}{\partial Y^{(a)}} = \partial_A \Lambda^A , \nn \\
\delta \cL_{red} &=& \epsilon^{(\alpha)}K^{(i)}_{(\alpha)}(X)\frac{\partial \cL_{red}}{\partial X^{(i)}}.
\eea
As could be noted, the reduced $\delta \cL$ coincides with $\delta\cL_{red}$ if $L^{(a)}_{(\alpha)}\rightarrow 0$ upon the reduction. On the other hand, as the six dimensional action is invariant, $\delta \cL = \partial_A \Lambda^A$  is the six dimensional divergence, and $\partial_A \Lambda^A\rightarrow \partial_\uA \Lambda^{\uA}$ if $\partial_4 = \partial_5 =0$. Then the variation of the four dimensional Lagrangian $\cL_{red}$ is the four dimensional divergence, and the corresponding action $S_{red} = \int d^4 x \cL_{red}$ is invariant.

We also need to make sure that the reduced transformations form a closed algebra. Calculating the commutators of the supersymmetries with the parameters $\epsilon_1{}^{(\alpha)}$, $\epsilon_2{}^{(\beta)}$ on $X$ and $Y$, we find
\bea\label{commXYtr}
\delta_1 \delta_2 X^{(i)}- \delta_2 \delta_1 X^{(i)} = -\epsilon_1{}^{(\alpha)}\epsilon_2{}^{(\beta)}\Big[ K^{(j)}_{(\alpha)}\frac{\partial K^{(i)}_{(\beta)}}{\partial X^{(j)}} + K^{(j)}_{(\beta)}\frac{\partial K^{(i)}_{(\alpha)}}{\partial X^{(j)}}  + L^{(a)}_{(\alpha)}\frac{\partial K^{(i)}_{(\beta)}}{\partial Y^{(a)}} + L^{(a)}_{(\beta)}\frac{\partial K^{(i)}_{(\alpha)}}{\partial Y^{(a)}} \Big], \nn \\
\delta_1 \delta_2 Y^{(a)}- \delta_2 \delta_1 Y^{(a)} = -\epsilon_1{}^{(\alpha)}\epsilon_2{}^{(\beta)}\Big[ L^{(b)}_{(\alpha)}\frac{\partial L^{(a)}_{(\beta)}}{\partial Y^{(b)}} + L^{(b)}_{(\beta)}\frac{\partial L^{(a)}_{(\alpha)}}{\partial Y^{(b)}}  + K^{(i)}_{(\alpha)}\frac{\partial L^{(a)}_{(\beta)}}{\partial X^{(i)}} + K^{(i)}_{(\beta)}\frac{\partial L^{(a)}_{(\alpha)}}{\partial X^{(i)}} \Big].
\eea
We assume that they commute properly, producing appropriate derivatives of $X$ and $Y$.

The reduced transformations and their commutator just read
\be\label{commXtr}
\delta_{red} X^{(i)} = \epsilon^{(\alpha)}K^{(i)}_{(\alpha)}(X), \;\; \delta_{red{}1} \delta_{red{}2} X^{(i)}- \delta_{red{}2} \delta_{red{}1} X^{(i)} = -\epsilon_1{}^{(\alpha)}\epsilon_2{}^{(\beta)}\Big[ K^{(j)}_{(\alpha)}\frac{\partial K^{(i)}_{(\beta)}}{\partial X^{(j)}} + K^{(j)}_{(\beta)}\frac{\partial K^{(i)}_{(\alpha)}}{\partial X^{(j)}} \Big].
\ee
Thus, to obtain \p{commXtr} by reducing \p{commXYtr}, we should assume that upon the reduction
\be\label{redcond}
Y^{(a)}\rightarrow 0, \;\; L^{(a)}_{(\alpha)} (X,Y) \rightarrow 0, \;\; \frac{\partial L^{(a)}_{(\beta)}}{\partial X^{(i)}} \rightarrow 0.
\ee
The latter two requirements could be satisfied if $L(X,Y)\sim Y$. If \p{redcond} hold, the commutator of the reduced transformations is equivalent to the reduced commutator of the original transformations, acting on $X$. It should be noted that the algebra of supersymmetries closes on the equations of motion and, therefore, the equations of motion that follow from the reduced Lagrangian should coincide with the result of reduction of the original equations of motion. However, this is true for arbitrary $\cL\big(\psi^\alpha_i,\partial_A \psi^\alpha_i, \cF_{AB}\big)$ and could not be considered as a limitation.

Now we should verify that the transformations \p{Sactvar}, \p{Qtranscomp} really satisfy the conditions \p{redcond}. As in our case the transformations are active and the variation of the derivative of a field is the derivative of the transformation of this field, the conditions \p{redcond} obviously hold for $\partial_4$, $\partial_5$ of any field. Thus $\cF_{45}$ could be the only obstacle. Let us consider this field in detail. It sufficient to study $F_{45}$, as $\cF_{45} = \cE_4{}^A \cE_5{}^B F_{AB} \rightarrow F_{45}$ upon the reduction.

As $\delta^\star_S F_{45} = H^C \partial_C F_{45}$, the broken supersymmetry is compatible with the reduction, and we need to consider the unbroken supersymmetry. Most convenient way to prove the fact that $\delta^\star_Q F_{45}=0$ upon the reduction is to use the following trick. Let us studyexpression \p{Qtranscomp} multiplied by $Z^{-1}$:
\be\label{deltaQVZinv}
\delta^\star_Q V_\alpha{}^\lambda \, \big( Z^{-1} \big)_\lambda{}^\beta = -4\im \epsilon^\mu_i \mD_{\mu\alpha}\psi^{i\nu}\big( Z^{-1} \big)_\nu{}^\beta-4\im \epsilon^\beta_i \mD_{\alpha\mu}\psi^{\nu i}  \big( Z^{-1} \big)_\nu{}^\mu + 2\im \delta_\alpha{}^\beta \epsilon^\gamma_i \mD_{\gamma\mu}\psi^{\nu i}\big( Z^{-1} \big)_\nu{}^\mu.
\ee
Its trace over $\alpha$, $\beta$ vanishes, as it should be. In the vector notation \p{VZinvvect} this implies the relation
\be\label{deltaQVZinvtrace}
\delta_Q^\star A = \frac{1}{2} G^{AB}\delta^\star_Q V_{AB}.
\ee
Multiplying \p{deltaQVZinv} by $\big(\gamma^{AB}\big)_\beta{}^{\alpha}$, one obtains another relation that is satisfied by the variation of $V_{AB}$
\bea\label{deltaQVZinvgammaAB}
&&\delta^\star_Q V_{AB} - \delta^\star_Q V_{AK}G_B{}^K + \delta^\star_Q V_{BK}G_A{}^K - \frac{1}{4}\epsilon_{ABCDKL}\delta^\star_Q V^{CD} G^{KL}+ \delta^\star_Q A\, G_{AB} = \nn \\
&&=-\im \big( \epsilon^\alpha_i \mD_A \psi^{\beta i} \big(\gamma_B\big)_{\alpha\beta} -  \epsilon^\alpha_i \mD_B \psi^{\beta i} \big(\gamma_A\big)_{\alpha\beta} \big) + \im \big( \epsilon^\alpha_i \mD_A \psi^{\beta i} G_{BC}\big(\gamma^C \big)_{\alpha\beta} -  \epsilon^\alpha_i \mD_B \psi^{\beta i} G_{AC}\big(\gamma^C\big)_{\alpha\beta} \big) -\\
&&-\frac{\im}{2} \big( \epsilon^\alpha_i \mD_A \psi^{\beta i} \big(\gamma_{BCD}\big)_{\alpha\beta} -  \epsilon^\alpha_i \mD_B \psi^{\beta i} \big(\gamma_{ACD}\big)_{\alpha\beta} \big)G^{CD}.\nn
\eea
One can note that the left-hand side can be represented as $\big(1 + \frac{3}{4}A^2 + \frac{1}{8}V_{KL}V^{KL} \big)^{-1}\Sigma_A{}^C \Sigma_B{}^D \delta^\star_Q F_{CD}$, where the matrix $\Sigma_A{}^B$ was defined in \p{Sigmamatr}. The simplest way to prove this is to use the relation
\bea\label{Sigma2deltaF}
\Sigma_A{}^C \Sigma_B{}^D \delta F_{CD} = \delta\big( \Sigma_A{}^C \Sigma_B{}^D  F_{CD} \big) -\delta \Sigma_A{}^C \Sigma_B{}^D  F_{CD}-\Sigma_A{}^C \delta\Sigma_B{}^D  F_{CD} = \nn \\
=\delta\Phi_{AB} + \big(\eta_{BC} A +V_{BC}\big)\delta\Sigma_A{}^C - \big(\eta_{AC}A +V_{AC}  \big)\delta\Sigma_B{}^C.
\eea
Therefore, multiplying \p{deltaQVZinvgammaAB} by two matrices $\big( \Sigma^{-1}  \big)_{A^\prime}{}^A \, \big( \Sigma^{-1}  \big)_{B^\prime}{}^B$, one can note that the variation of $F_{AB}$ consists only of terms that contain $\partial_{A^\prime} \psi^{\nu i}$ or $\partial_{B^\prime} \psi^{\nu i}$. As a result, the variation of $F_{45}$ under the unbroken supersymmetry is proportional to the fields that vanish upon the reduction, just as needed. It can be stated in other words: the Bianchi identity \p{BIvect} and the unbroken supersymmetry transformations are compatible with each other.

The above results guarantee that the reduced action would be invariant with respect to reduced broken and unbroken supersymmetries, and the reduction of the action and the Bianchi identity can now be performed explicitly.

The six dimensional spinor $\psi^{\alpha}_i$ can be represented as the doublet
\be\label{psired}
\psi^\alpha_i = \sqrt{2}\left( \begin{array}{l}
\psi^{\ualpha}_i   \\
\bpsi^{\dot\ualpha}_i
\end{array}   \right).
\ee
Then the elements of the fermionic matrix reduce to
\bea\label{cEred}
&&\ucE_{\uA}{}^{\uB} = \delta_{\uA}{}^{\uB} + \im \big( \bpsi^{\dot\ualpha i} \partial_{\uA} \psi^{\ualpha}_i + \psi^{\ualpha}_i \partial_{\uA} \bpsi^{\dot\ualpha i} \big) \big( \sigma^{\uB} \big)_{\ualpha\dot{\ualpha}}, \nn \\
&&\cE_{\uA}{}^{4}+\im \cE_{\uA}{}^{5} = 2\im \psi^i_{\ualpha}\partial_{\uA}\psi^{\ualpha}_i, \;\; \cE_{\uA}{}^{4}-\im \cE_{\uA}{}^{5} = 2\im \bpsi^i_{\dot\ualpha}\partial_{\uA}\bpsi^{\dot\ualpha}_i, \;\; \cE_{4}{}^{A} =\delta_{4}^A, \;\; \cE_5{}^A = \delta_5^A.
\eea
The broken supersymmetry covariant derivative should be defined as $\ucD_\uA = \big(\ucE^{-1}\big)_{\uA}{}^\uB \partial_{\uB}$.

If $\partial_4 = \partial_5=0$ and $\cF_{45}=0$, the Bianchi identity $\partial_{[C}\cF_{AB]}=0$ implies that the four dimensional vectors $\cF_{\uA 4 }$, $\cF_{\uA 5}$ are the derivatives of some scalar fields. We express the derivative of the complex scalar field and its conjugate by the relations
\bea\label{FWred}
\partial_{\uA} W = \frac{\im}{2}\big( {\cal F}_{\uA 4} +\im {\cal F}_{\uA 5}  \big), \;\; \partial_{\uA} \bW = -\frac{\im}{2}\big( {\cal F}_{\uA 4} -\im {\cal F}_{\uA 5}  \big) \nn\\
{\cal F}_{\uA 4} =\cE_{\uA}{}^{D}F_{D4} = \cE_{\uA}{}^{\uD}F_{\uD 4}, \;\; {\cal F}_{\uA 5} =\cE_{\uA}{}^{D}F_{D5} = \cE_{\uA}{}^{\uD}F_{\uD 5} \; \Rightarrow \\
F_{\uA 4} = -\im \big({\underline{\cal D}}_{\uA} W - {\underline{\cal D}}_{\uA} \bW \big), \;\; F_{\uA 5} = - \big({\underline{\cal D}}_{\uA} W + {\underline{\cal D}}_{\uA} \bW\big).
\eea
The reduced four dimensional field strength tensor, including fermionic contributions, reads
\be\label{Fred}
{\cal F}_{\uA\uB} = \ucE_{\uA}{}^{\uC}\ucE_{\uB}{}^{\uD}F_{\uC\uD} +2 \big( \partial_\uA W \bpsi^i_{\dot\ualpha}\partial_{\uB}\bpsi^{\dot\ualpha}_i - \partial_\uB W\bpsi^i_{\dot\ualpha}\partial_{\uA}\bpsi^{\dot\ualpha}_i    \big)-2 \big( \partial_\uA \bW \psi^i_{\ualpha}\partial_{\uB}\psi^{\ualpha}_i -\partial_\uB \bW  \psi^i_{\ualpha}\partial_{\uA}\psi^{\ualpha}_i  \big).
\ee
It satisfies the Bianchi identity $\partial_{[\uC}\cF_{\uA\uB]}=0$.

Finally, the reduced Lagrangian reads
\bea\label{Lagrred}
{\cal L}_{red} &=& -\det\ucE {\cal L}_{main} -2 \im \widetilde{\cal F}{}^{\uA\uB} \big(\partial_\uA W \bpsi^i_{\dot\ualpha}\partial_{\uB}\bpsi^{\dot\ualpha}_i   + \partial_\uA \bW  \psi^i_{\ualpha}\partial_{\uB}\psi^{\ualpha}_i   \big)-\nn \\
&&-2\epsilon^{\uA\uB\uC\uD}\big( \bpsi^{\dot\ualpha i} \partial_{\uA} \psi^{\ualpha}_i + \psi^{\ualpha}_i \partial_{\uA} \bpsi^{\dot\ualpha i} \big) \big( \sigma_{\uB} \big)_{\ualpha\dot{\ualpha}}\partial_\uC W \partial_\uD \bW , \\
&\mbox{where}& {\cal L}_{main} =1 + \sqrt{-\det\big( \eta_{\uA\uB} + F_{\uA\uB} - 2 \ucD_{\uA} W  \ucD_{\uB} \bW -  2\ucD_{\uB} W  \ucD_{\uA} \bW \big)}. \nn
\eea
Note that in the limit $F_{\uA\uB}\rightarrow 0$, $\psi^\ualpha_2 \rightarrow 0$, $\bpsi^{\dot\ualpha 2}\rightarrow 0$, this Lagrangian coincides with the Lagrangian of the 3-brane in $D=6$ \cite{scalaracts}, up to the redefinition $x^\uA \rightarrow - x^\uA$. Its invariance with respect to the broken and unbroken supersymmetries is guaranteed by the previous considerations, and we do not need to check it explicitly.

Let us now discuss the self-duality properties of the Lagrangian \p{Lagrred}. The concept of self-duality can be introduced in standard nonlinear four dimensional electrodynamics, starting from the fact that in such theories the Bianchi identity and the equation of motion of the abelian gauge field have the same structural form:
\be\label{calGdef}
\partial_\uB {\widetilde\cF}{}^{\uA\uB}=0, \; \; \partial_\uB {\widetilde {\cG}}^{\uA\uB} =0, \;\; {\widetilde {\cG}}^{\uA\uB} = 2 \frac{\partial {\cL} \big( \cF \big)}{\partial {\cF}_{\uA\uB}}.
\ee
Then one may consider the $U(1)$ transformations, which preserve the set of Bianchi identities and the equations of motion:
\be\label{sdtr}
\cF^\prime_{\uA\uB} = \cos(\lambda)\cF_{\uA\uB}+ \sin(\lambda)\cG_{\uA\uB}, \;\; \cG^\prime_{\uA\uB} = \cos(\lambda)\cG_{\uA\uB}- \sin(\lambda)\cF_{\uA\uB}
\ee
If $\cL\big( \cF^\prime \big) = \cL\big( \cF\big)$, the model is called self-dual. As a consequence, it is also self-dual with respect to Legendre transformations. It was proved in \cite{nlselfdualprev} that a theory of a single abelian field is self-dual if the following equation is satisfied:
\be\label{sdcond1}
\epsilon_{\uA\uB\uC\uD}{\widetilde {\cal G}}^{\uA\uB} {\widetilde {\cal G}}^{\uC\uD} - \epsilon^{\uA\uB\uC\uD}{\cal F}_{\uA\uB}{\cal F}_{\uC\uD}=0.
\ee
The Lagrangian of the standard bosonic Born-Infeld theory is a solution to this equation. Moreover, it is worth to noting that \p{sdcond1} is also satisfied by the Lagrangian of the component $N=2$ supersymmetric Born-Infeld theory \cite{compN2d4BI} and by the bosonic core of the $N=4$ theory \p{Lagrred}. One may go further and calculate ${\widetilde{\cal G}}$, which corresponds to the full Lagrangian \p{Lagrred}:
\be\label{tildeG}
{\widetilde {\cal G}}^{\uA\uB} = -2\det\ucE \frac{\partial{\cal L}_{main}}{\partial F_{\uC\uD}}\big(\cE^{-1}\big)_\uC{}^\uA \big( \cE^{-1}  \big)_\uD{}^\uB +2\im \epsilon^{\uA\uB\uC\uD}\big( \bpsi^i_{\dot\ualpha}\partial_{\uC}\bpsi^{\dot\ualpha}_i \partial_\uD W + \psi^i_{\ualpha}\partial_{\uC}\psi^{\ualpha}_i \partial_\uD \bW  \big).
\ee
Substituting it into equation \p{sdcond1}, one can find that it is not satisfied, however. After some algebra, the result of the substitution can be represented as
\bea\label{tildeG2}
\det\ucE \Big(4 \epsilon_{\uA\uB\uC\uD}\frac{\partial{\cal L}_{main}}{\partial F_{\uA\uB}}\frac{\partial{\cal L}_{main}}{\partial F_{\uC\uD}} - \epsilon^{\uA\uB\uC\uD}F_{\uA\uB}F_{\uC\uD}\Big)+ \nn \\
-32 \im \det\ucE \frac{\partial{\cal L}_{main}}{\partial F_{\uA\uB}} \big( \cD_\uA W  \bpsi^i_{\dot\ualpha}\cD_{\uB}\bpsi^{\dot\ualpha}_i + \cD_\uA \bW \psi^i_{\ualpha}\cD_{\uB}\psi^{\ualpha}_i \big)-\\- 8\epsilon^{\uA\uB\uC\uD}\cF_{\uA\uB}\big( \partial_\uC W \bpsi^i_{\dot\ualpha}\partial_{\uD}\bpsi^{\dot\ualpha}_i - \partial_\uC \bW \psi^i_{\ualpha}\partial_{\uD}\psi^{\ualpha}_i \big).\nn
\eea
The first line of this expression should vanish if the bosonic core of the action is self-dual, which is exactly the case. The second line does not vanish, however. This means that the Lagrangian \p{Lagrred} could be self-dual only if the scalars and the fermions are assumed to transform nontrivially. Indeed, one can adapt the methods of \cite{nlselfdualprev} to find that if $W$, $\psi_i^\ualpha$ are not inert, the following equation should be satisfied:
\be\label{sdcond2}
\delta_{W,\psi} L = \frac{\lambda}{8} \big( \epsilon_{\uA\uB\uC\uD}{\widetilde {\cal G}}^{\uA\uB} {\widetilde {\cal G}}^{\uC\uD} - \epsilon^{\uA\uB\uC\uD}{\cal F}_{\uA\uB}{\cal F}_{\uC\uD}  \big),
\ee
where $\delta_{W,\psi} L$ is the variation of the Lagrangian, when $W$, $\psi_i^\ualpha$ but not $\cF_{\uA\uB}$ are varied. Taking into account that the transformations of $W$, $\psi^\alpha_i$ also induce the nontrivial transformation of $F_{\uA\uB} = F_{\uA\uB}\big(W,\psi,\cF  \big)$, one can find the following very simple variations of $W$, $\psi^\ualpha_i$ which allow equation \p{sdcond2} to be satisfied:
\be\label{sdcond3}
\delta W = \im \lambda W, \;\; \delta \bW = -\im\lambda\bW, \;\; \delta \psi^\ualpha_i = \im \lambda \psi_i^\ualpha, \;\; \delta \bpsi^{i \dot\ualpha} = -\im \lambda \bpsi^{i \dot\ualpha}.
\ee
These phase rotations keep the matrix $\cE_\uA{}^\uB$, as well as the covariant derivatives, inert. Actually, the only terms in the Lagrangian which nontrivially transform are the first Wess-Zumino term, containing ${\widetilde\cF}_{\uA\uB}$, and $\cL_{main}$, purely due to its dependence on $F_{\uA\uB}$.

Finally, let us compare this result with the others previously obtained using the superfield methods.

In \cite{nlselfdual}, the concept of dualities was generalized to explicitly supersymmetric theories and the appropriate conditions of the theory to be self-dual were derived in the cases of $N=1$ and $N=2$ supersymmetry. It was again found that the $N=1$ theory satisfies this new superfield condition, and it was suggested that explicitly $N=2$ supersymmetric generalization of the Born-Infeld theory also should be self-dual. The duality transformations and the self-duality condition in this case were found to be
\bea\label{N2sdcond}
&\delta \boldsymbol{\cal W} = \lambda \boldsymbol{\cal M}, \;\; \delta\boldsymbol{\cal M} = -\lambda \boldsymbol{\cal W}, \;\;  \boldsymbol{\cal M} = -\im \frac{\delta}{\delta \boldsymbol{\cal W}}S\big[\boldsymbol{\cal W},{\overline {\boldsymbol{\cal W}}}\big], \;\; {\overline {\boldsymbol{\cal M}}} = \im \frac{\delta}{\delta {\overline {\boldsymbol{\cal W}}}}S\big[\boldsymbol{\cal W},{\overline {\boldsymbol{\cal W}}}\big],& \nn \\
&\int d^4 x d^4\theta \big( {\boldsymbol{\cal M}}^2 + {\boldsymbol{\cal W}}^2 \big) = \int  d^4 x d^4\bar\theta  \big( {\overline {\boldsymbol{\cal M}}}^2 + {\overline {\boldsymbol{\cal W}}}^2 \big).&
\eea
The self-duality condition is satisfied by the superfield action,
\bea\label{SFact}
S = \frac{1}{2}\int d^4 x \big( D^4 {\boldsymbol{\cal W}}^2 + \bD{}^4 {\overline {{\boldsymbol{\cal W}}}}{}^2 + D^4 \bD{}^4 \big( {\boldsymbol{\cal W}}^2  {\overline {{\boldsymbol{\cal W}}}}{}^2 \big)\big) + \ldots, \;\; \Rightarrow \\
\boldsymbol{\cal M} =-\im \boldsymbol{\cal W} \big( 1+ \bD{}^4  {\overline {{\boldsymbol{\cal W}}}}{}^2  \big)+ \ldots\;\; \overline{\boldsymbol{\cal M}} =-\im \overline{\boldsymbol{\cal W}} \big( 1+ D{}^4   {{\boldsymbol{\cal W}}}{}^2  \big)+\ldots
\eea
where we write down only the terms up to fourth power in the fields. In \cite{nlselfdual}, terms up to ${\boldsymbol{\cal W}}^8$ were obtained and they coincided with the terms obtained in \cite{sfN4BI1} from the condition of the invariance of the action with respect to the spontaneously broken supersymmetry.

The $N=2$ vector multiplet itself is defined by the irreducibility conditions
\be\label{cWirr}
\bD_{\dot\ualpha i}{\boldsymbol{\cal W}} =0 \;\; D^i_{\ualpha }{\overline {{\boldsymbol{\cal W}}}} =0, \;\; D^\ualpha_i D_{\ualpha j}{\boldsymbol{\cal W}} = \bD_{\dot\ualpha i} \bD^{\dot\ualpha }_j {\overline {{\boldsymbol{\cal W}}}}.
\ee
Note that while obtaining the variation of the action with respect to the ${\boldsymbol{\cal W}}$, ${\overline {{\boldsymbol{\cal W}}}}$, only chirality conditions should be taken into account, but not the Bianchi identity.

The components of the superfields ${\boldsymbol{\cal W}}$, ${\overline {{\boldsymbol{\cal W}}}}$  can be defined as
\bea\label{calWcomps}
&\cW={\boldsymbol{\cal W}}|_{\theta\rightarrow 0},\;\; \cbW ={\overline {{\boldsymbol{\cal W}}}}|_{\theta\rightarrow 0}, \;\; \xi^{\ualpha}_i = \frac{\im}{2}D^\ualpha_i {\boldsymbol{\cal W}}|_{\theta\rightarrow 0}, \;\;  {\bar\xi}^{\dot\ualpha i} =\frac{\im}{2}\bD^{\dot\ualpha i}{\overline {{\boldsymbol{\cal W}}}}|_{\theta\rightarrow 0},\;\; {\cal B}^{ij} =  D^{i\ualpha}D_\ualpha^j  {\boldsymbol{\cal W}}|_{\theta\rightarrow 0},& \nn \\
& D^{i\ualpha}D_i{}^{\ubeta}{\boldsymbol{\cal W}}|_{\theta\rightarrow 0} = \im \big(\sigma^{\uA\uB}   \big)^{\ualpha\ubeta}\cH_{\uA\uB}, \;\; \bD^{i\dot\ualpha}\bD_i{}^{\dot\ubeta}{\overline {{\boldsymbol{\cal W}}}}|_{\theta\rightarrow 0} = \im \big(\tilde\sigma^{\uA\uB}   \big)^{\dot\ualpha\dot\ubeta}\cH_{\uA\uB}, \;\; \partial_{[\uC}\cH_{\uA\uB]}=0.&
\eea

We are primarily interested in terms up to the fourth power in bosons and of the second power in bosons and fermions. Calculating the component form of the action \p{SFact}, one may note that the auxiliary field should be defined as
\be\label{SFactaux}
B^{ij} ={\cal B}^{ij} + \im\big( \xi^{(i\ualpha}{\bar\xi}_k^{\dot\ualpha} + \xi^\ualpha_k {\bar\xi}^{\dot\ualpha (i} \big)\partial_\uA {\cal B}^{j)k} \big( \sigma^\uA   \big)_{\ualpha\dot\ualpha} - \cW \cbW \partial^2 {\cal B}^{ij} + \ldots
\ee
Then its equation of motion becomes algebraic and can be easily solved. The solution for $B^{ij}$ is at least cubic in the fields and, therefore, contributes only to the sixth order terms in the Lagrangian. Therefore, $B^{ij}$ can be safely dropped. After that, using the following redefinitions of the basic fields,
\bea\label{redefcomps}
W &=& \cW \Big( 1 - \frac{1}{8}\cH_{\uA\uB}\cH^{\uA\uB}  +\frac{\im}{8}\cH_{\uA\uB}{\widetilde{\cH}}^{\uA\uB}  \Big) - \partial_\uA \cW\, \partial^\uA \cW \, \cbW - \cW\cbW \partial^2 \cW +\nn \\&& +\im\partial_\uA \cW \big( \sigma^{\uA} \big)_{\ualpha\dot{\ualpha}} \xi^{\ualpha}_i \, {\bar\xi}^{i \dot\ualpha}-2\im\cW \big( \sigma^{\uA} \big)_{\ualpha\dot{\ualpha}} {\bar\xi}^{i \dot\ualpha}\partial_\uA  \xi^{\ualpha}_i + \ldots , \nn \\
\psi^{\ualpha}_i &=& \xi^{\ualpha}_i \Big( 1 - \frac{1}{8}\cH_{\uA\uB}\cH^{\uA\uB}  +\frac{\im}{8}\cH_{\uA\uB}{\widetilde{\cH}}^{\uA\uB} - \frac{1}{4}\big( \cW \partial^2{\cbW} + \cbW \partial^2 \cW  \big)  \Big) +\nn \\ &&+ \frac{\im}{2} \cH_{\uA\uB} \big(\tilde\sigma^{\uB}  \big)^{\dot\ualpha\ualpha}\partial^{\uA}\big( \cW{\bar\xi}_{i\dot\ualpha} \big)+
+\frac{1}{2}{\widetilde\cH}^{\uA\uB}\big(\tilde\sigma_{\uB}  \big)^{\dot\ualpha\ualpha}\big( \partial_\uA \cW\, {\bar\xi}_{i\dot\ualpha} -  \cW\,\partial_\uA {\bar\xi}_{i\dot\ualpha} \big)- \\&& - \cbW \partial_\uB \cW  \partial^\uB \xi^\ualpha_i - \frac{1}{2}\cbW \partial^2 \big( \cW \xi^\ualpha_i \big) - \cW \partial^\uA \big( \cbW \partial_\uA \xi^\alpha_i  \big) - \im \cW \partial_\uA \cbW \partial_\uB \xi^\ubeta_i \big( \sigma^{\uA\uB} \big)_\ubeta{}^\ualpha + \ldots,\nn \\
{\cal F}_{\uA\uB} &=& \frac{1}{2} \cH_{\uA\uB} + \partial_\uA \partial^\uD \big( \cW\cbW \cH_{\uB\uD}  \big) + \frac{\im}{4}\partial_\uA \big( \widetilde\cH_{\uB\uD} \big(\cbW \partial^\uD \cW - \cW \partial^\uD \cbW \big) \big) -\nn  \\&&- \partial_\uA \big(\widetilde\cH_{\uB\uD}\big( \sigma^\uD \big)_{\ualpha\dot\ualpha}\xi^\ualpha_i {\bar\xi}^{\dot\ualpha i}   \big)
-2\im \partial_\uA \big( \cW \partial^\uC \big({\bar\xi}^{\dot\ualpha i} {\bar\xi}^{\dot\ubeta}_{i} \big)  \big)\big(\tilde\sigma_{\uB\uC}   \big)_{\dot\ualpha\dot\ubeta}-2\im \partial_\uA \big( \cbW \partial^\uC \big({\xi}^{\ualpha i} {\xi}^{\ubeta}_{i} \big)  \big)\big(\sigma_{\uB\uC}   \big)_{\ualpha\ubeta}-\nn  \\&&-2 \partial_\uA \big( \cW {\bar\xi}^{\dot\ualpha}_i \partial_\uB {\bar\xi}_{\dot\ualpha}^i\big)+2\partial_\uA \big( \cbW {\xi}^{\ualpha}_i \partial_\uB \xi_{\ualpha}^i\big)+ \ldots - \big( \uA \leftrightarrow \uB \big), \nn
\eea
one can obtain the Lagrangian
\bea\label{SFactcomp}
{\cal L}_{SF} &=& 2 \partial_\uA W\, \partial^\uA \bW - \frac{1}{4}{\cal F}_{\uA\uB}\, {\cal F}^{\uA\uB} -2\im \big(\psi^\ualpha_i \partial_\uA \bpsi^{\dot\ualpha i} +\bpsi^{\dot\ualpha i}  \partial_\uA \psi^\ualpha_i \big)\big( \sigma^\uA \big)_{\ualpha  \dot\ualpha}+ \nn \\
&& + 2 \partial_\uA W \partial^\uA W\, \partial_\uB \bW \partial^\uB \bW- 2 {\cal F}_\uA{}^\uC {\cal F}^{\uA\uD}\partial_\uC W \, \partial_\uD \bW + \frac{1}{2} {\cal F}_{\uA\uB}\, {\cal F}^{\uA\uB} \partial^\uC W \partial_\uC \bW + \nn\\
&&- \frac{\im}{2}  \big(\psi^\ualpha_i \partial_\uA \bpsi^{\dot\ualpha i} +\bpsi^{\dot\ualpha i}  \partial_\uA \psi^\ualpha_i \big)\big( \sigma^\uA \big)_{\ualpha  \dot\ualpha} \Big(\partial^\uC W \partial_\uC \bW  - \frac{1}{8}{\cal F}_{\uC\uD}\, {\cal F}^{\uC\uD}   \Big) +  \\
&& + \frac{\im}{2} \big(\psi^\ualpha_i \partial^\uA \bpsi^{\dot\ualpha i} +\bpsi^{\dot\ualpha i}  \partial^\uA \psi^\ualpha_i \big)\big( \sigma^\uB \big)_{\ualpha  \dot\ualpha} \big( \partial_\uA W \partial_\uB \bW + \partial_\uB W \partial_\uA \bW - \frac{1}{2}{\cal F}_{\uA\uC}{\cal F}_{\uB}{}^{\uC}  \big) - \nn \\
&& -2 {\cal F}^{\uA \uB} \big( \partial_\uB W \bpsi^{\dot\ualpha}_i \partial_\uA \bpsi^i_{\dot\ualpha} -\partial_\uB \bW \psi^{\ualpha}_i \partial_\uA \psi^i_{\ualpha} \big) +2\im \widetilde {\cal F}^{\uA\uB}\big( \bW \partial_\uA \psi^\ualpha_i \partial_\uB \psi_\ualpha^i + W \partial_\uA \bpsi^{\dot\ualpha}_i \partial_\uB \bpsi^i_{\dot\ualpha}  \big) + \nn \\
&& -2 \epsilon^{\uA\uB\uC\uD}\big( \bpsi^{\dot\ualpha i} \partial_{\uA} \psi^{\ualpha}_i + \psi^{\ualpha}_i \partial_{\uA} \bpsi^{\dot\ualpha i} \big) \big( \sigma_{\uB} \big)_{\ualpha\dot{\ualpha}}\partial_\uC W \partial_\uD \bW + \ldots\nn
\eea
In the considered approximation, it coincides with the power expansion of \p{Lagrred}.

Let us note that the redefinition of the components is somewhat arbitrary. For example, the term
\be
2\im \cbW \partial^2 \cW \xi^{\ualpha }_i \partial_{\uA}\bxi^{\dot\ualpha i} \big( \sigma^A \big)_{\ualpha\dot\ualpha} \nn
\ee
can be absorbed into $\bW$, $\bW = \cbW - \im \cbW  \xi^{\ualpha }_i \partial_{\uA}\bxi^{\dot\ualpha i} \big( \sigma^A \big)_{\ualpha\dot\ualpha} + \ldots$ but also can be used to redefine the fermion $\psi^\ualpha_i = \xi_i^\ualpha + 2 \cbW \partial^2 \cW \xi^{\ualpha }_i + \ldots$. To properly resolve this arbitrariness, one should take into account that $W$, $\psi^\ualpha_i$, $\bpsi^{\dot\ualpha i}$, ${\cal F}_{\uA\uB}$ are the components of some superfield $\mW$, satisfying some nonlinear covariant chirality constraint. While we do not consider such a superfield, it is worth noting that in \cite{compN4d4BIold} the approximate relation between the superfields satisfying standard linear and covariant nonlinear constraints was obtained and the first component of this relation is exactly the definition of the proper scalar component $W$ \p{redefcomps}.

It is worthwhile to note, however, that the duality transformations \p{sdtr}, \p{sdcond3} do not coincide with the component expansion of the transformations \p{N2sdcond}. Indeed, using the relation
\be\label{sigmarel}
\big(\sigma^{\uA\uB}\big)_{\alpha}{}^\beta \big(\sigma^{\uC\uD}\big)_{\beta}{}^\alpha = 2 \big(\eta^{\uA\uC}\eta^{\uB\uD}-\eta^{\uA\uD}\eta^{\uB\uC} + \im \epsilon^{\uA\uB\uC\uD}   \big),
\ee
one can establish that in the lowest approximation
\be\label{sfcomptr}
\delta \cW = -\im \lambda \cW + \ldots, \;\; \delta \xi^\ualpha_i =-\im \lambda \xi^\ualpha_i, \;\; \delta \cH_{\uA\uB} =\lambda {\widetilde{\cH}}_{\uA\uB}+\ldots,
\ee
with some signs being opposite of \p{sdtr}, \p{sdcond3}. Moreover, these transformations also contain higher derivative terms and they do not vanish even after passing to the components of nonlinear realization. For example, the variation of $W$ \p{redefcomps} reads
\be\label{anotherWvar}
\delta W = -\im \lambda W -4\lambda W \bpsi^{\dot\ualpha i} \partial_A \psi^{\ualpha}_i \big( \sigma^A \big)_{\alpha\dot\alpha} +2\im \lambda W\bW\partial^2 W+\ldots
\ee
The relation between these two sets of transformations, therefore, remains unclear.

As the last comment, it is desirable to compare the action \p{Lagrred} with the component actions previously obtained in the paper \cite{Kallosh}. Though the notations and the general form of the Lagrangians are significantly different, one can expect that the Lagrangian of the $N=4$, $d=4$ Born-Infeld theory, obtained in \cite{Kallosh}, is related to \p{Lagrred} by a field redefinition. To shed some light on this point, one may study the broken supersymmetry transformation laws. Initially, it seems natural to associate the parameter $\zeta$ (formula (A.30) in \cite{Kallosh}) with the standard broken supersymmetry transformations, as the transformation law of the fermion with the parameter $\zeta$ is just the law of Volkov and Akulov. The bosonic fields, however, are not covariant with respect to the $\zeta$ transformations. In particular, the transformation laws of the scalar fields begin with the terms proportional to the transformation parameter and the fermionic field. The only appropriate Lorentz and $SU(2)$ covariant terms are $\delta\phi \sim \zeta^\ualpha_i \psi^i_\ualpha$ and $\delta \bar\phi \sim \bar\zeta^{\dot\ualpha}_i \bpsi^i_{\dot\ualpha}$. It is natural to expect that $\phi$, $\bar\phi$ differ from $W$, $\bW$ by terms with these transformation laws. The expected terms have to be quadratic in fermions, but the only candidates $\psi^{i\ualpha}\psi_{i\ualpha}$ and $\bpsi^{i\dot\ualpha}\bpsi_{i\dot\ualpha}$ are equal to zero identically. The only solution to this problem seems to identify the standard broken supersymmetry transformations with $\delta_\zeta + \delta_{\epsilon}|_{\epsilon \rightarrow \zeta}$. This identification removes the undesirable terms in the transformation laws of the bosons but introduces terms $\sim \bar\zeta \beta \lambda $ in the transformation law of the Goldstone fermions. Then, to remove these newly appeared terms, one should perform the redefinition of the fermionic field $\psi\sim\lambda+{\cal F}\Gamma \lambda +\ldots$. Note that in the Lagrangian this redefinition generates the terms linear in $\cal F$ and quadratic in fermions, which are absent in \p{Lagrred} but typically occur in the actions in the paper \cite{Kallosh}. Therefore, a tentative conclusion could be reached that from the point of view of the standard nonlinear realizations the action obtained in \cite{Kallosh} corresponds to the breaking of $Q+S$ supersymmetry. The exact correspondence between the action \cite{Kallosh} and \p{Lagrred} should be studied elsewhere.

\section*{Conclusion}
In this article, the $N=(2,0)$, $d=6$ Born-Infeld theory was considered in the component approach. It was shown that it is possible to construct its component action using the principles already successfully employed in the construction of the component $N=2$, $d=4$ Born-Infeld theory \cite{compN2d4BI}. These include the use of the standard nonlinear realization formalism with the exponential parametrization of the coset space to find the transformation laws of the superfields with respect to both unbroken and spontaneously broken supersymmetries and automorphisms, as well as the differential forms and the derivatives covariant with respect to these transformations. Another important idea used in this paper, already employed in \cite{compN2d4BI}, is that the properly generalized irreducibility conditions of the vector multiplets should be invariant not only with respect to broken supersymmetry but also with respect to the subgroup of the external automorphisms of the supersymmetry algebra. With these ideas implemented, it becomes a difficult though technical problem to calculate the Bianchi identity, which is satisfied by the bosonic field strength, and prove its equivalence to the standard one. The fermionic contributions to the identity can be unambiguously restored by demanding its covariance with respect to broken supersymmetry. The rest of the procedure is common to all studied component actions with partial spontaneous breaking of supersymmetry. It involves modifying the bosonic action following the recipe of Volkov and Akulov \cite{neugold}, adding the Wess-Zumino term, and checking the invariance with respect to unbroken supersymmetry, fixing the remaining arbitrary constants in the process.

The reduction of the constructed theory to four dimensions was also considered. It was proven that the supersymmetry transformations do not contradict the reduction conditions and, therefore, the action after reduction is still invariant with respect to the $N=4$, $d=4$ supersymmetry. Its self-duality was proven at the component level, with rather simple duality transformations of the scalar and the fermionic fields. The comparison with the previous works shows that the obtained action coincides with ones found in \cite{nlselfdual}, \cite{sfN4BI1} in the second and the fourth power in the fields after the proper field redefinition. Whether the action obtained in \cite{Kallosh} is nontrivially different from \p{Lagrred} remains unclear.

Let us also mention the observation made during the analysis of the bosonic Bianchi identity for the field strength. This identity involves the matrix which, at the same time, relates the anticommutator of two spinor derivatives to the $x^A$ derivative, relates the physical bosonic field strength to the tensor component of the multiplet, and is used to multiply the original identity to bring it to the proper form. Therefore, the role of this matrix is likely fundamental for the component $D$-brane actions and requires further investigation.

\section*{Acknowledgements}
The work of N.K. was supported by the RFBR, grant 18-52-05002 Arm\_a. The author wishes to thank Stefano Bellucci, Sergey Krivonos, Anton Sutulin and Armen Yeranyan who collaborated with him on the component approach to partial breaking of global supersymmetry.

\section*{Appendix: Properties of the $\gamma^A$-matrices}
The six dimensional $\gamma^A$-matrices are assumed to have following properties \cite{HST6dsusy}:
\bea\label{gammadef}
&&\big(\gamma{}^A\big)_{\alpha\beta} = -\big(\gamma{}^A\big)_{\beta\alpha}, \;\; \big(\gamma{}^A\big)_{\alpha\beta}\big(\gamma_A \big)_{\mu\nu} = -2 \epsilon_{\alpha\beta\mu\nu}, \;\; \big(\tilde{\gamma}{}^A\big)^{\alpha\beta} = \frac{1}{2}\epsilon^{\alpha\beta\mu\nu}\big(\gamma{}^A\big)_{\mu\nu},
 \nn \\
&&\epsilon^{\alpha\beta\mu\nu}\epsilon_{\alpha\beta\mu\nu}=24,\;\;\big( \gamma{}^A   \big)_{\alpha\lambda} \big(\tilde{\gamma}{}^{B}\big)^{\lambda\beta} +  \big( \gamma{}^B   \big)_{\alpha\lambda} \big(\tilde{\gamma}{}^{A}\big)^{\lambda\beta} = 2\eta^{AB}\delta_\alpha{}^\beta.
\eea

The composite matrices are defined as:
\bea\label{gamma23def}
&&\big(\gamma{}^{AB}\big)_\alpha{}^\beta = \frac{1}{2}\Big(\big( \gamma{}^A   \big)_{\alpha\lambda} \big(\tilde{\gamma}{}^{B}\big)^{\lambda\beta} -  \big( \gamma{}^B   \big)_{\alpha\lambda} \big(\tilde{\gamma}{}^{A}\big)^{\lambda\beta} \Big), \nn \\
&&\big(\gamma^{ABC}\big)_{\alpha\beta} = \frac{1}{2}\Big(\big(\gamma^A\big)_{\alpha\rho}\big(\tilde\gamma^B\big)^{\rho\sigma}\big(\gamma^C \big)_{\sigma\beta} + \big(\gamma^A\big)_{\beta\rho}\big(\tilde\gamma^B\big)^{\rho\sigma}\big(\gamma^C \big)_{\sigma\alpha}\Big), \nn\\
&&\big(\tilde{\gamma}{}^{ABC}\big)^{\alpha\beta} = \frac{1}{2}\Big(\big(\tilde\gamma{}^A\big)^{\alpha\rho}\big(\gamma{}^B\big)_{\rho\sigma}\big(\gamma^C \big)^{\sigma\beta} + \big(\tilde\gamma{}^A\big)^{\beta\rho}\big(\gamma{}^B\big)_{\rho\sigma}\big(\tilde\gamma{}^C \big)^{\sigma\alpha}\Big).
\eea
The composite matrices satisfy the relations
\bea\label{gamma23prop}
&&\big(\gamma{}^{AB} \big)_\alpha{}^\beta = -\big(\gamma{}^{BA} \big)_\alpha{}^\beta, \;\; \big(\gamma{}^{AB} \big)_\alpha{}^\alpha =0,\nn \\
&&\big(\gamma{}^{AB} \big)_\alpha{}^\lambda \big(\gamma{}^{CD} \big)_\lambda{}^\beta = -\Big( \eta^{AC}\big(\gamma{}^{BD} \big)_\alpha{}^\beta - \eta^{BC}\big(\gamma{}^{AD} \big)_\alpha{}^\beta -\eta^{AD}\big(\gamma{}^{BC} \big)_\alpha{}^\beta + \eta^{BD}\big(\gamma{}^{AC} \big)_\alpha{}^\beta\Big) -\nn \\
&&- \frac{1}{2}\epsilon^{ABCDMN}\big(\gamma_{MN} \big)_\alpha{}^\beta - \big(\eta^{AC}\eta^{BD}-\eta^{BC}\eta^{AD}\big)\delta_\alpha{}^{\beta},  \\
&&\big(\gamma{}^{ABC}\big)_{\alpha\beta} = \big(\gamma{}^{ABC}\big)_{\beta\alpha} = \big(\gamma{}^{[ABC]}\big)_{\alpha\beta}, \;\;  \big(\tilde\gamma{}^{ABC}\big)^{\alpha\beta} = \big(\tilde\gamma{}^{ABC}\big)^{\beta\alpha} = \big(\tilde\gamma{}^{[ABC]}\big)^{\alpha\beta}, \nn \\
&&\epsilon_{ABCMNP} \big(\gamma{}^{MNP}\big)_{\alpha\beta} = -6 \big(\gamma{}_{ABC}\big)_{\alpha\beta}, \;\; \epsilon_{ABCMNP} \big(\tilde\gamma{}^{MNP}\big)^{\alpha\beta} = 6 \big(\tilde\gamma{}_{ABC}\big)^{\alpha\beta},\nn \\
&&\big(\gamma{}^{ABC}\big)_{\alpha\beta} \big(\tilde\gamma{}^{MNP}\big)^{\alpha\beta} =-4 \big( \eta^{AM}\eta^{BN}\eta^{CP}- \eta^{AN}\eta^{BM}\eta^{CP}-\eta^{AM}\eta^{BP}\eta^{CN} - \eta^{AP}\eta^{BN}\eta^{CM} +\nn \\&&+ \eta^{AP}\eta^{BM}\eta^{CN}+ \eta^{AN}\eta^{BP}\eta^{CM}  \big) +4 \epsilon^{ABCMNP}.\nn
\eea

The vector and spinor notation for the vectors and antisymmetric tensors are related in the following way:
\be\label{vectspnot}
P_{\alpha\beta} = \frac{1}{2}\big(\gamma^A \big)_{\alpha\beta}P_A \;\; \Leftrightarrow \;\; P_A = -\frac{1}{2}\big(\tilde\gamma^A \big)^{\alpha\beta}P_{\alpha\beta}, \;\;
F_{\alpha}{}^{\beta} = \frac{1}{2}\big(\gamma^{AB} \big)_{\alpha}{}^{\beta}F_{AB} \;\; \Leftrightarrow \;\; F_{AB} = - \frac{1}{4}\big(\gamma_{AB} \big)_{\alpha}{}^{\beta}F_{\beta}{}^\alpha.
\ee

For the purposes of dimensional reduction, one can use the following explicit representation of these matrices: \bea\label{gammarepr}
\big(\gamma^{\uA}\big)_{\alpha\beta} = \left(\begin{array}{cc}
0 & \big(\sigma^\uA\big)_{\ualpha\dot\ualpha}  \\
-\big(\sigma^\uA\big)_{\ualpha\dot\ualpha} & 0
\end{array}  \right), \;\; \big(\gamma^{4}\big)_{\alpha\beta} = \left(\begin{array}{cc}
\epsilon_{\ualpha\ubeta} & 0  \\
0 & \epsilon_{\dot\ualpha\dot\ubeta}
\end{array}  \right), \;\; \big(\gamma^{5}\big)_{\alpha\beta} = \left(\begin{array}{cc}
-\im \epsilon_{\ualpha\ubeta} & 0  \\
0 & \im \epsilon_{\dot\ualpha\dot\ubeta}
\end{array}  \right).
\eea
Here, $\sigma^\uA = \big( {\mathbf 1}, \sigma^1, \sigma^2, \sigma^3   \big)$ are standard four dimensional $\sigma$-matrices, while the $\epsilon$-tensors have properties
\bea\label{eps2dprop}
\epsilon_{\ualpha\ubeta} = -\epsilon_{\ubeta\ualpha}, \;\; \epsilon_{\ualpha\ubeta}\epsilon^{\ubeta\ulambda} = \delta_\ualpha^\ulambda, \;\; \epsilon_{12}=1, \nn \\
\epsilon_{\dot\ualpha\dot\ubeta} = -\epsilon_{\dot\ubeta\dot\ualpha}, \;\; \epsilon_{\dot\ualpha\dot\ubeta}\epsilon^{\dot\ubeta\dot\ulambda} = \delta_{\dot\ualpha}^{\dot\ulambda}, \;\; \epsilon_{\dot{1} \dot{2}}=1.
\eea
 They can be used to lower four dimensional spinor indices, $\psi_{\ualpha i} = \epsilon_{\ualpha\ubeta}\psi^{\ubeta}_i$, e.t.c.

\end{document}